\documentclass[sigconf]{acmart}
\usepackage{etoolbox}
\usepackage{mathrsfs}
\usepackage{mathtools}
\usepackage{multirow}
\usepackage{graphicx}

\usepackage{booktabs} 
\usepackage{subcaption}
\usepackage{balance}
\usepackage{bbm}
\usepackage{amsmath}
\usepackage{amsfonts}
\usepackage{tabulary}

\newcommand{\RN}[1]{%
	\textup{\lowercase\expandafter{\it \romannumeral#1}}%
}

\AtBeginDocument{%
  \providecommand\BibTeX{{%
    \normalfont B\kern-0.5em{\scshape i\kern-0.25em b}\kern-0.8em\TeX}}}

\copyrightyear{2021}
\acmYear{2021} 
\setcopyright{iw3c2w3}
\acmConference[WWW '21]{Proceedings of the Web Conference 2021}{April 19--23, 2021}{Ljubljana, Slovenia} 
\acmBooktitle{Proceedings of the Web Conference 2021 (WWW '21), April 19--23, 2021, Ljubljana, Slovenia}
\acmPrice{}
\acmDOI{10.1145/3442381.3450110}
\acmISBN{978-1-4503-8312-7/21/04}

\begin{document}

\title{Leveraging User Behavior History for Personalized Email Search}

\author{Keping Bi$^1$, Pavel Metrikov$^{2,4}$, Chunyuan Li$^3$, Byungki Byun$^2$}
\affiliation{
  $^1$ University of Massachusetts Amherst, $^2$ Microsoft, $^3$ Microsoft Research, \\ $^4$ Institute for Information Transmission Problems, Russian Academy of Sciences
  \\
\texttt{kbi@cs.umass.edu, \{pametrik, chunyl, bybyun\}@microsoft.com}
}

\begin{abstract}

An effective email search engine can facilitate users' search tasks and improve their communication efficiency. Users could have varied preferences on various ranking signals of an email, such as relevance and recency based on their tasks at hand and even their jobs. Thus a uniform matching pattern is not optimal for all users. Instead, an effective email ranker should conduct personalized ranking by taking users' characteristics into account. Existing studies have explored user characteristics from various angles to make email search results personalized. However, little attention has been given to users' search history for characterizing users. Although users' historical behaviors have been shown to be beneficial as context in Web search, their effect in email search has not been studied and remains unknown. Given these observations, we propose to leverage user search history as query context to characterize users and build a context-aware ranking model for email search. In contrast to previous context-dependent ranking techniques that are based on raw texts, we use ranking features in the search history. This frees us from potential privacy leakage while giving a better generalization power to unseen users. 
Accordingly, we propose a {\em context-dependent neural ranking model} (CNRM) that encodes the ranking features in users' search history as query context and show that it can significantly outperform the baseline neural model without using the context. We also investigate the benefit of the query context vectors obtained from CNRM on the state-of-the-art learning-to-rank model LambdaMart by clustering the vectors and incorporating the cluster information. Experimental results show that significantly better results can be achieved on LambdaMart as well, indicating that the query clusters can characterize different users and effectively turn the ranking model personalized.

\end{abstract}


\keywords{Email Search, Query Context, Personalized Search}

\maketitle
\section{Introduction}
\label{sec:introduction}

Email has long been an indispensable way for both business and personal communication in our daily life. As email storage quotas become large due to cheap data storage, users do not feel the need to delete emails and let their mailboxes keep growing \cite{carmel2015rank,mackenzie2019exploring}. Therefore, an effective email search engine is quite important to support users' search needs and facilitate users' daily life.

In email search, users could have varied preferences on email's various ranking signals such as relevance, recency, and other email properties due to their primary purposes of the accounts and their occupation. For instance, commercial users (the employees of a company that uses the email service) and consumer users (individual email service users) could have significantly different preferred email properties, such as the length of the email threads and the number of recipients and attachments. 
While it can be known explicitly whether a user is commercial or consumer, other implicit user cohorts also exist for whom different aspects of emails matter more for ranking.
For example, in contrast to engineers, it is more likely that the emails salespeople want to re-find have them as senders rather than as recipients. Users who organize their mailboxes more often could prefer recency than relevance since they can search inside a specific folder that already satisfies some filtering conditions. Hence, it could improve search quality by identifying diversified implicit user cohorts as query context and conducting personalized ranking accordingly.

Existing work has explored various types of user information to provide personalized email search results. \citet{zamani2017situational} leverage the situational context of a query such as geographical (e.g., country, language) and temporal (e.g., weekday, hour) features to characterize users. 
\citet{weerkamp2009using} used users' email threads and mailing lists as contextual information to expand queries. \citet{kuzi2017query} trained local word embeddings with the content of each user's mailbox and used the nearest neighbors of the query terms in the embedding space as the query context. 
\citet{shen2018multi} collected the frequent n-grams of the results retrieved from the user's mailbox with a decent initial ranker, based on which the query is clustered, and the cluster information can be considered as context for conducting adaptive ranking. 

Despite the extensive studies on exploring user characteristics for personalized email search, little attention has been paid to characterizing users with their search history. Leveraging users' historical click-through data as context has been shown to be beneficial in Web search \cite{shen2005implicit,liao2013vlhmm, ustinovskiy2013personalization, white2010predicting, xiang2010context,croft2005context,bennett2012modeling, matthijs2011personalizing}. 
However, the effect of incorporating users' historical behaviors in email search has not been studied and remains unknown. 
There are substantial differences between email search and Web search. To name a few, first, email search can only be conducted on users' personal corpus instead of a shared public corpus, and the target is usually to re-find a known item; second, email content cannot be exposed to third parties due to privacy concern, which makes it hard to investigate and analyze ranking models; third, in email search logs, cross interactions with the same item from different users do not exist and search queries may be hard to generalize across users since they are personal and pertain to their particular email content. 
Due to the differences mentioned above, it is valuable to investigate the benefit of using user search history as context for email search, and it is also challenging to incorporate this information effectively.

In this paper, we study the effect of characterizing users with their search history as query context in personalized email search. As far as we know, this is the first work in this direction. We construct the context from user history grounded on ranking features instead of raw texts as previous context-aware ranking models do. These numerical features free us from potential privacy leakage present in a neural model trained with raw texts while having better generalization capabilities to unseen users. They can also capture users' characteristics from a different perspective compared with their semantic intents indicated by the terms or n-grams in their click-through data.
Based on both neural models and the prevailing state-of-the-art learning-to-rank model - LambdaMart \cite{burges2010ranknet}, we investigate the query context obtained from the ranking features in the user search history. Specifically, we propose a neural baseline ranking model based on the numerical ranking features and a context-dependent neural ranking model (CNRM) that encodes the ranking features of users' historical queries and their associated clicked documents as query context. We further cluster each query context vector learned from CNRM to reveal hidden user cohorts that are captured in users' search behaviors. To examine the effectiveness of using the learned user cohorts, we then incorporate the cluster information into the LambdaMart model. Experimental results show that the query context helps improve the ranking performance on both neural models and LambdaMart models. 

Our contributions in this paper can be summarized as follows:
1) we propose a context-dependent neural ranking model (CNRM) to incorporate the query context encoded from numerical features extracted from users' search history to characterize potential different matching patterns for diversified user groups;
2) we conduct a thorough evaluation of CNRM on the data constructed from the search logs of one of the world's largest email search engines and show that CNRM can outperform the baseline neural model without context significantly;
3) we cluster the query context vectors and incorporate the query cluster information with LambdaMart and produce significantly better ranking performances;
4) we analyze the information carried in the query context vectors, the feature weight distribution of the query clusters in the LambdaMart model, and users' distribution in terms of their query clusters. 
\section{Related Work}
\label{sec:related_work}
Our paper is directly related to two threads of work: email search and context-aware ranking in web search. 

\textbf{Email Search.}
\label{subsec:emailsearch}
There have been limited existing studies on email search, probably due to the lack of publicly available email data, which is too private and sensitive to expose. The TREC Enterprise track dataset \cite{craswell2005overview, soboroff2006overview} is an exception. The dataset contains about 200k email messages crawled from the W3C public mailing list archive \footnote{lists.w3.org} and 150 <query, document-id> pairs for email re-finding tasks. Earlier studies on email search are mainly based on this dataset.  \citet{macdonald2006combining} proposed a field-based weighting model that can combine Web features with email features and investigated where field evidence should be used or not. \citet{craswell2005microsoft} scored email messages based on their multi-field with the BM25f formula \cite{robertson2004simple}. \citet{ogilvie2005experiments} combined evidence from the subject, content of an email, the thread the email occurs in, and the content in the emails that are in reply to the email with a language model approach. \citet{yahyaei2008applying} conducted maximum entropy modeling to estimate feature weights in known-item email retrieval. \citet{weerkamp2009using} studied to improve email search quality by expanding queries with information sources such as threads and mailing lists. 

Some research also uses two other datasets to study email search which are Enron \footnote{\url{http://www.cs.cmu.edu/~enron/}} and PALIN \footnote{\url{https://ropercenter.cornell.edu/ipoll/study/31086601}} \cite{abdelrahman2010new,bhole2015correcting}. Enron is an enterprise email corpus that the Federal Energy Regulatory Commission released after the Enron investigation. PALIN is the personal email corpus of the former Alaskan governor Sarah Palin. Both of them do not have queries or relevance judgments of emails. \citet{abdelrahman2010new} generated 300 synthetic queries for Enron using topic words related to Enron and estimate email relevance by three judges according to given guidelines. \citet{bhole2015correcting} conducted misspelling query correction on queries constructed by volunteers for both Enron and PALIN. Similar to the W3C TREC dataset, they are far from real scenarios as the queries, and the relevance judgments do not originate from the mailbox owners. 

In recent years, more studies have been conducted with real queries and judgments derived from users' interactions with the results in the context of large Web email services. 
Some work has studied the effect of displaying results ranked based on relevance or recency \cite{carmel2015rank, ramarao2016inlook, carmel2017promoting}.
\citet{carmel2015rank} followed a two-phase ranking scheme where in the first stage recall is emphasized based on exact or partial term matching, and in the second stage, a learning-to-rank (LTR) model is used to balance the relevance and recency for ranking. This way outperforms the widely adopted time-based ranking by a large margin. Aware that users still prefer chronological order and not all of them accept ranking by relevance,
\citet{ramarao2016inlook} designed a system called InLook that has an intent panel that shows three results based on relevance as well as a recency panel that shows the rest results based on recency. They further studied the benefit of each part of the system. \citet{carmel2017promoting} proposed to combine h relevance results and k-h time-sorted results in a fixed display window of size k. Then they investigated the performances of allowing or forbidding duplication in the two sets of results and using a fixed or variable h. \citet{mackenzie2019exploring} observed that time-based ranking begins to fail as email age increases, suggesting that hybrid ranking approaches with both relevance-based rankers and the traditional time-based rankers could be a better choice. Nowadays, most email service providers have adopted the strategy of two separate panels for relevance and recency based results and allowing duplicates. 

Efforts have also been made to improve the search quality of relevance-based results for the relevance panel.
\citet{aberdeen2010learning} exploited social features that capture the degree of interaction between the sender and the recipient for search in Gmail Priority Inbox. \citet{bendersky2017learning} proposed an attribution parameterization method that uses the click-through rate of frequent n-grams of queries and documents for email ranking. \citet{kuzi2017query} studied query expansion methods for email search by using a translation model based on clicked messages, finding neighbor terms based on word embedding models trained with each user's mailbox, and extracting terms from top-ranked results in the initial retrieval with pseudo relevance feedback techniques. \citet{li2019multi} proposed a multi-view embedding based synonym extraction method to expand queries for email search. \citet{zamani2017situational} incorporated the situational context such as geographical (country, language) and temporal (weekday, hour) features to facilitate personal search. \citet{shen2018multi} clustered queries based on the frequent n-grams of results retrieved with a baseline ranker and then used the query cluster information as an auxiliary objective in multi-task learning. There has also been research that studies how to combine sparse and dense features in a unified model effectively \cite{meng2020separate}, combine textual information from queries and documents with other side information to conduct effective and efficient learning to rank \cite{qin2020matching}, and transfer models learned from one domain to another in email search \cite{tran2019domain}.

To the best of our knowledge, there has been no previous work on characterizing different users' preferred ranking signals with users' search history as query context and letting the ranking model adapt to the context, especially based on numerical features rather than terms or n-grams of the queries and email documents as most previous work uses. 

\textbf{Context-aware Ranking in Web Search.}
\label{subsec:context-rank}
Context-aware ranking has been explored extensively in Web search, which is usually based on users' long or short-term search history \cite{shen2005implicit,liao2013vlhmm, ustinovskiy2013personalization, white2010predicting, xiang2010context,croft2005context,bennett2012modeling, matthijs2011personalizing,lobel2019ract}. Short-term search history includes the queries issued in the current search session and their associated clicked results. \citet{shen2005implicit} proposed a context-aware language modeling approach that can extract expansion terms based on short-term search history and showed the effectiveness of the model on TREC collections. Afterward, some studies also focused on leveraging users' short-term history as context, especially their clicked documents \cite{liao2013vlhmm, ustinovskiy2013personalization, white2010predicting, xiang2010context}. Long-term search history is not limited to the search behaviors in the current session and can include more users' historical information. It is often used for search personalization \cite{croft2005context,bennett2012modeling, matthijs2011personalizing}. 

Although we also leverage users' search history to characterize users, we focus on differentiating the importance of various ranking signals for different users rather than refining the semantic intent of a user query with long or short-term context as in Web search. More importantly, we aim to study the effectiveness of search history in the context of email search, which is significantly different from Web search in terms of privacy concerns and the property of search logs. 

\section{Method}
\label{sec:method}
In this section, we first describe the problem formulation and available ranking features. Then we propose two neural models, a) a  vanilla neural ranking model without using context (NRM) and b) a context-dependent neural ranking model (CNRM) that encodes users' historical behaviors as query context. Afterward, we introduce how we optimize these neural models. At last, we illustrate how to incorporate the query context information with LambdarMart models.

\begin{table}
	\caption{Representative Features in Personal Email Search. Note that no word or n-gram information is available. }
	\centering
	\label{tab:feature}
	\small
	\setlength\extrarowheight{3pt}
	\scalebox{0.95}{    
		\begin{tabular}{p{1.7cm}  p{1.0cm}  p{0.8cm} l}
			\hline
            Feature Group & Type & Notation & Examples \\
            \hline
            \multirow{2}{*}{query-level} & discrete & $F^{\mathcal{D}}_Q$ & query language, user type \\
            & continuous & $F^{\mathcal{C}}_Q$ & IDF of query terms \\
            \multirow{2}{*}{document-level} & discrete & $F^{\mathcal{D}}_D$ & flagged, read, meta-field length \\
            & continuous & $F^{\mathcal{C}}_D$ & recency, email length\\
            q-d matching & continuous & $F^{\mathcal{C}}_{QD}$ & BM25, language model scores \\
			\hline
		\end{tabular}
	}
\end{table}

\begin{figure*}
	\centering
	\includegraphics[width=1.0\textwidth]{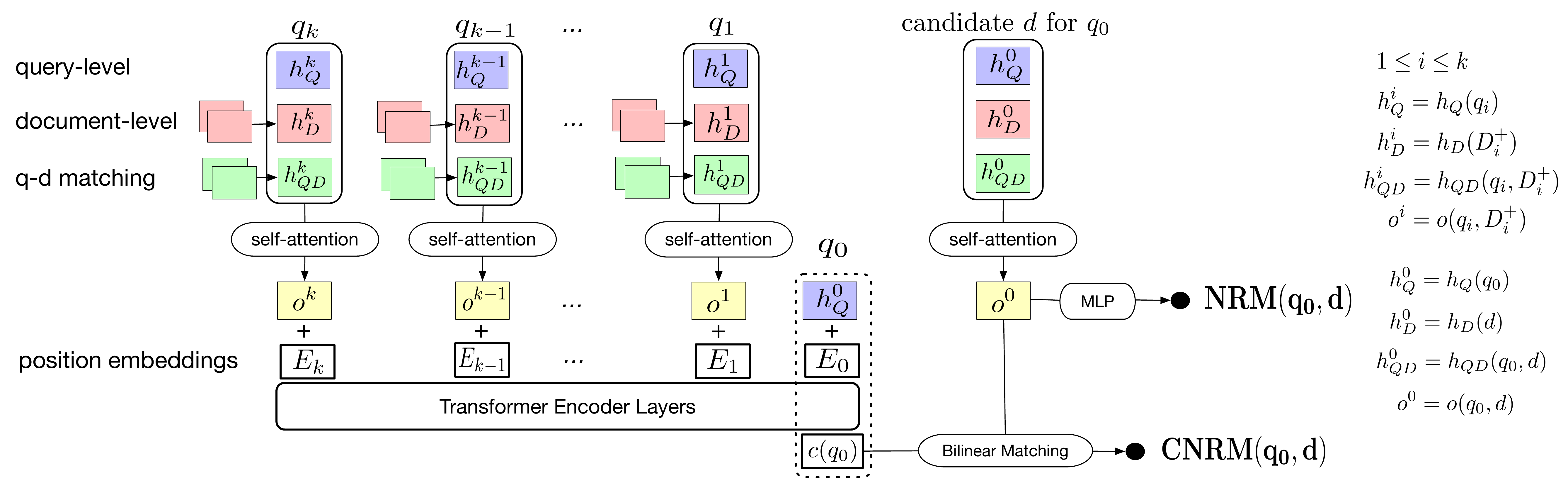} 
	\caption{The model architecture of the vanilla neural ranking model (NRM) and the context-dependent neural ranking model (CNRM). NRM only uses the features of a candidate $d$ given $q_0$ to compute its ranking score, while CNRM considers extra context information encoded from the features corresponding to $[q_0,...,q_k]$ and their associated clicked documents.}
	\label{fig:model}
\end{figure*}

\subsection{Background and Problem Formulation}
\label{subsec:background}
Let $q$ be a query issued by user $u_{q}$, $D$ be the set of candidate documents \footnote{We will use document and email interchangeably in this paper. } for $q$ which can be obtained from the retrieval model at an early stage. $D$ consists of user $u_{q}$'s clicked documents $D^+$ and the other competing documents $D^-$, i.e., $D=D^+ \cup D^-$. The objective of email ranking is to promote $D^+$ to the top so that users can find their target documents as soon as possible. 

{\bf Email Ranking Features.}~
For any document $d\in D$, three groups of features are extracted from a specifically designed privacy-preserved system. Table \ref{tab:feature} shows some representative features of each group. 
$(\RN{1})$
{\em Query-level features} $F_Q(q)$ indicate properties of a user's query and the user's mailbox, such as a user type representing whether $u_q$ is a consumer (general public) or a commercial user (employee of a commercial company to which the email service is provided). $F_Q$ can be further divided to discrete features $F^{\mathcal{D}}_Q$ and continuous features $F^{\mathcal{C}}_Q$, i.e., $F_Q(q) = F^{\mathcal{D}}_Q(q) \cup F^{\mathcal{C}}_Q(q)$.
$(\RN{2})$
{\em Document-level features} $F_D(d)$ characterize properties of $d$ such as its recency and size, independent of $q$. $F_D$ also consists of discrete $F^{\mathcal{D}}_D$ and continuous features $F^{\mathcal{C}}_D$, i.e., $F_D(d) = F^{\mathcal{D}}_D(d) \cup F^{\mathcal{C}}_D(d)$. 
$(\RN{3})$
{\em query-document (q-d) matching features} $F_{QD}(q,d)$ measure how $d$ matches the query $q$ in terms of overall and each field in an email, such as ``to'' or ``cc''list, and subject. The $q$-$d$ matching features $F_{QD}(q,d)$ are all considered as continuous features, i.e., $F_{QD}(q,d) = F^{\mathcal{C}}_{QD}(q,d)$. The score of $d$ given $q$ can be computed based on $F_{Q}(q), F_{D}(d)$, and $F_{QD}(q,d)$. To protect user privacy, raw text of queries and documents are not available during our investigation. 

{\bf User Search History.}~
When we further consider user behavior history for the current query $q_0$ issued by $u_{q_0}$, more information is available to rank documents in the candidate set $D_0$. Let $(q_k, q_{k-1}, \cdots, q_1)$ be the recent $k$ queries issued by $u_{q_0}$ before $q_0$ and $(D^+_k, D^+_{k-1}, \cdots, D^+_{1})$ be their associated relevant documents. The features of the user's historical positive q-d pairs are used to identify the user's matching patterns and facilitate the ranking for $q_0$.


\subsection{Vanilla Neural Ranking Model}
\label{subsec:vanilla_nrm}
We first describe a baseline method: the vanilla neural ranking model (NRM) only uses the three available groups of ranking features of the current query-document (q-d) pair for ranking without considering user search history. NRM first encodes each group of features to the latent space, then aggregates the hidden vectors of each group of features with a self-attention mechanism. At last, it scores the q-d pair based on the aggregated representation. The model architecture can be found in Figure \ref{fig:model}. We will introduce each component of the model next. 

\textbf{Feature Encoding.}
The three groups of email features have different scales; thus it is important to normalize and map them to the same latent space to maximize the information the model can learn. Discrete features, such as user type, query language, and whether the email has be read, will be projected to hidden vectors by referring to the corresponding embedding lookup tables. Continuous features such as email length and recency will be considered as numerical input vectors directly. 
Specifically, take query-level features for instance, discrete features $F^{\mathcal{D}}_Q(q)$ will be mapped to
\begin{equation}
\label{eq:qdiscrete1}
    g^{\mathcal{D}}_Q(q) = [emb(f_1)^T; \cdots; emb(f_{|F^{\mathcal{D}}_Q(q)|})^T]^T
\end{equation}
where $[\cdot;\cdot;\cdot]$ denotes the concatenation of a list of vectors, $|\cdot|$ indicates the dimension of a vector, and $emb(f)\in \mathbb{R}^{e \times 1}$ is the embedding of feature $f$ with dimension $e$. Then the concatenated vector is mapped to $h^{\mathcal{D}}_Q \in \mathbb{R}^{m \times 1}$ with matrix $W^{\mathcal{D}}_Q \in \mathbb{R}^{m \times |g^{\mathcal{D}}_Q(q)|}$, where $m$ is the dimension of the hidden vector $h^{\mathcal{D}}_Q$, according to
\begin{equation}
\label{eq:qdiscrete2}
    h^{\mathcal{D}}_Q(q) = \tanh(W^{\mathcal{D}}_Q g^{\mathcal{D}}_Q(q)).
\end{equation}

Similarly, the continuous query-level features $F^{\mathcal{C}}_Q(q)$ are mapped to $h^{\mathcal{C}}_Q \in \mathbb{R}^{m \times 1}$ with matrix $W^{\mathcal{C}}_q \in \mathbb{R}^{m \times |F^{\mathcal{C}}_Q(q)|}$, i.e., 
\begin{equation}
\label{eq:qcont}
    h^{\mathcal{C}}_Q(q) = \tanh(W^{\mathcal{C}}_Q g^{\mathcal{C}}_Q(q))
\end{equation}
Then the two hidden vectors are concatenated as the representation of query-level features of $q$, denoted as $h_Q(q)\in \mathbb{R}^{2m \times 1}$, namely,
\begin{equation}
\label{eq:qoverall}
    h_Q(q) = [h^{\mathcal{D}}_Q(q)^T; h^{\mathcal{C}}_Q(q)^T]^T
\end{equation}
Likewise, the hidden vector corresponding to the document-level features $F_D(d)$ can be obtained as $h_D(d) = [h^{\mathcal{D}}_D(d)^T; h^{\mathcal{C}}_D(d)^T]^T \in \mathbb{R}^{2m \times 1}$ following the same encoding paradigm but with different lookup tables for discrete features and mapping matrices in the above equations. 

Since all the q-d matching features $F_{QD}(q,d)$ are continuous features, they are mapped to
\begin{equation}
\label{eq:qd_cont}
h_{QD}(q,d) = h^{\mathcal{C}}_{QD}(q,d) = \tanh(W^{\mathcal{C}}_{QD} F_{QD}(q,d))
\end{equation}
where $W^{\mathcal{C}}_{QD} \in \mathbb{R}^{2m \times |F_{QD}(q,d)|}$.
Note that the bias terms in all the mapping functions throughout the paper are omitted for simplicity. 

\textbf{Feature Aggregation with Self-attention}. To better balance the importance of each group of features in the ranking process, the hidden vectors corresponding to the query, document, and q-d matching features are aggregated according to a self-attention mechanism as follows. \footnote{We also tried to aggregate the vectors of the document and q-d matching features using their attention weights with respect to the vector of the query features. However, this way is not better than using self-attention so we do not include this in the paper.} They are first mapped to another vector with $W_a\in \mathbb{R}^{2m \times 2m}$:
\begin{equation}
\label{eq:qdqdmap}
\begin{split}
    & o_Q(q) = \tanh(W_{a}h_{Q}(q)); \\
    & o_D(d) = \tanh(W_{a}h_{D}(d)); \\
    & o_{QD}(q,d) = \tanh(W_{a}h_{QD}(q,d))
\end{split}
\end{equation}
$o_Q(q), o_D(d)$, and $o_{QD}(q,d)$ are all in $\mathbb{R}^{2m \times 1}$. 
Then the attention weight of each component is computed based on softmax function of its dot-product with itself and the other two components. The new vector of each component is the weighted combination of the input and average pooling is applied to the new vectors to obtain the final vector. In concrete, the aggregated vector $o(q,d) \in \mathbb{R}^{2m \times 1}$ is computed according to
\begin{equation}
\label{eq:attn_agg}
\begin{split}
    & o_c = [o_Q(q);o_D(d);o_{QD}(q,d)] \\
    & W_{attn} = \text{softmax}(o_{c}^T o_{c}) \\
    & o(q,d) = \text{AvgPool}(o_{c} W_{attn}) \\
\end{split}
\end{equation}

\textbf{Scoring Function.}
The score of document $d$ given $q$ is finally computed as:
\begin{equation}
\label{eq:nrm}
    NRM(q,d) = W_{s2}\tanh(W_{s1} o(q,d))
\end{equation}
where $W_{s1} \in \mathbb{R}^{m \times 2m}$ and $W_{s2} \in \mathbb{}{R}^{1 \times m}$.

\subsection{Context-dependent Neural Ranking Model}
\label{subsec:context_nrm}
A user's historical search behaviors could help characterize the user and provide a query-level context to the current query. We propose a context-dependent neural ranking model (CNRM) to extract a context vector based on the user's search history and scores a candidate document according to both its features given the current query and the query context. We first introduce how CNRM encodes a candidate document's features and then show how CNRM obtains the context vector and scores the candidate document. The model architecture is shown in Figure \ref{fig:model}.

Given the current query $q_0$, for each $d$ in the candidate set $D$, CNRM obtains the final hidden vector $o(q_0,d)$ from its initial three groups of features following the same paradigm in Section \ref{subsec:vanilla_nrm}. 

\textbf{Context Encoding.}
As shown in Figure \ref{fig:model}, for each historical query $q_i (1 \leq i \leq k)$, we obtain the hidden vector of its query-level features, i.e., $h_Q(q_i)$ using the same encoding process as the current query $q_0$ according to Equation \eqref{eq:qdiscrete1}, \eqref{eq:qdiscrete2}, and \eqref{eq:qcont}. Similarly, for each document $d_i$ in the relevant document set $D^+_i$ of $q_i$, we encode its document-level features and q-d matching features to the latent space using the same mapping functions and parameters that are used for the candidate document $d$ given $q_0$, which can be denoted as $h_D(d_i)$ and $h_{QD}(q_i, d_i)$ respectively. Then the aggregated document-level and q-d matching vectors corresponding to $D^+_i$ are computed as the projected average embeddings of each $d_i$ in $D^+_i$:
\begin{equation}
\begin{split}
    h_D(D^+_i) &= \tanh(W_{d} \text{Avg}(\{h_D(d_i)|d_i\in D^+_i\})) \\
    h_{QD}(q_i, D^+_i) &= \tanh(W_{qd} \text{Avg}(\{h_{QD}(q_i,d_i)|d_i\in D^+_i\})),
\end{split}
\end{equation}
where $W_d \in \mathbb{R}^{2m \times 2m}$ and $W_{qd} \in \mathbb{R}^{2m \times 2m}$. The overall vector corresponding to $q_i$, i.e., $o(q_i, D^+_i)$, is computed according to Equation \eqref{eq:qdqdmap} and \eqref{eq:attn_agg} with $h_D(d)$ and $h_{QD}(q,d)$ replaced by $h_D(D^+_i)$ and $h_{QD}(q_i, D^+_i)$ respectively.

To produce a context vector for $q_0$, we encode the sequence of the overall vectors corresponding to each historical query and the associated positive documents together with the hidden vector of the query-level feature of $q_0$ with a transformer \cite{vaswani2017attention} architecture. Specifically,
\begin{equation}
\label{eq:context}
\begin{split}
    c(q_0) = & \text{Transformer}(o(q_k,D^+_k)+posEmb(k), \cdots, \\
            &o(q_1,D^+_1)+posEmb(1), h_Q(q_0)+posEmb(0))
\end{split}
\end{equation}
As in Figure \ref{fig:model}, the output of Transformer~\cite{vaswani2017attention} is the vector corresponding to $q_0$ in the final transformer layer, which acts as the final context vector. \footnote{We do not introduce how Transformers work due to the space limitations. Please refer to \cite{vaswani2017attention} for details.} As such, the interaction between the current query-level features and the historical behaviors can be captured and used to better balance the importance of each search behavior in the history.

\textbf{Scoring Function.}
To allow each dimension of the encoded features of a candidate $d$ and the context vector interact sufficiently, the final score of $d$ given $q_0$ is computed by bilinear matching between the aggregated vector of $d$ and the encoded context:
\begin{equation}
\label{eq:cnrm}
    CNRM(q_0,d) = o(q_0,d)^T W_b c(q_0)
\end{equation}
where $W_b \in \mathbb{R}^{2m \times 2m}$, $o(q_0,d)$ and $c(q_0)$ are computed according to Equation \eqref{eq:qoverall} and \eqref{eq:context} respectively.

\subsection{Model Optimization}
\label{subsec:unbiased}
We use the softmax cross-entropy list-wise loss \cite{ai2018learning} to train NRM and CNRM. Specifically, the cross-entropy between the list of document labels $y(d)$ and the list of probabilities of each document obtained with the softmax function applied on their ranking scores are used as the loss function. For a query $q$ with candidate set $D$, the ranking loss $\mathcal{L_r}$ can be computed as:
\begin{equation}
\label{eq:original_loss}
    \mathcal{L_r} = -\sum_{d \in D} y(d) \log \frac{\exp(s_r(d))}{\sum_{d'\in D} \exp(s_r(d'))},
\end{equation}
where $s_r$ is a scoring function for $NRM$ in Equation \eqref{eq:nrm} or $CNRM$ and  in Equation \eqref{eq:cnrm}. 
However, the document label $y(d)$ is extracted based on user clicks and thus biased towards top documents since users usually examine the results from top to bottom and decide whether to click. To counteract the effect of position bias, better balance the feature weights, and let context take a properer effect, we train NRM and CNRM through unbiased learning. In contrast to Web search, where there is one panel for organic results, email search usually has two panels with several results ranked by relevance followed by results ranked by time. As stated in \cite{carmel2017promoting}, allowing duplicates in the two panels leads to better user satisfaction. So each document $d$ can have two positions in terms of the relevance and time window respectively, denoted as $pos_r(d)$, $pos_t(d)$. We adapt the unbiased learning algorithm proposed by \citet{ai2018unbiased} to our email search scenario. An examination model is built to estimate the propensity of different positions and adjust each document's weight in the ranking loss with inverse propensity weighting. Meanwhile, the ranking scores can adjust the document weights in the examination loss by inverse relevance weighting as well \cite{ai2018unbiased}. We will introduce the examination model and the collaborative model optimization next.

\textbf{Examination Model.} Two lookup tables are created for relevance and time positions respectively. Let $emb(pos_{r}(d)) \in \mathbb{R}^{m \times 1}$ and $emb(pos_{t}(d)) \in \mathbb{R}^{m \times 1}$ be the embeddings of relevance and time positions, $W_{p1}\in \mathbb{R}^{m \times 2m}$ and $W_{p2}\in \mathbb{R}^{1 \times m}$ be the mapping matrices. The score of $d$, i.e., $s_e(d)$, output by the examination model is computed as:
\begin{equation}
\begin{split}
    h_{p1}(d) &= [emb(pos_r(d))^T; emb(pos_t(d)^T]^T \\
    s_e(d) &= W_{p2}\tanh(W_{p1}h_1) \\
\end{split}
\end{equation}

\textbf{Collaborative Optimization}.
\label{subsec:optimation}
We follow the dual learning algorithm in \cite{ai2018unbiased} and use the softmax cross-entropy list-wise loss \cite{ai2018learning} to train both the examination model and the ranking model. Let $d_{r1}$ be the document that is ranked to the first position in the relevance panel, namely, $pos_r(d_{r1}) = 1$. Inverse propensity weighting based on the relative propensity compared with the first result is applied to each document and adjust the final loss. Then Equation \eqref{eq:original_loss} can be refined as:
\begin{equation}
\label{eq:loss}
\begin{split}
    g_e(d) &= \frac{\exp(s_e(d))}{\sum_{d'\in D} \exp(s_e(d'))} \\
    \mathcal{L}_r &= -\sum_{d \in D} y(d) \frac{g_e(d_{r1})}{g_e(d)} \log \frac{\exp(s_r(d))}{\sum_{d'\in D} \exp(s_r(d'))}
\end{split}
\end{equation}
Regularization terms in Equation \eqref{eq:loss} has been omitted for simplicity. Similarly, the loss of the examination model can be derived by swapping $s_r$ and $s_e$ in Equation \eqref{eq:loss}. The training of the examination and ranking models depends on each other and both of them are refined step by step until the models converge. 

\subsection{Incorporate Context to LambdaMart}
\label{subsec:context2lambdamart}
It is hard to directly incorporate user behavior history into some learning-to-rank (LTR) models, such as the state-of-the-art method -- LambdaMart \cite{burges2010ranknet}, since it is grounded on various ready-to-use features and does not support learning features automatically during training. To study whether user search history can characterize the context of a query $q$ and benefit search quality for LambdaMart, we investigate a method to incorporate the context vector encoded from user search history, i.e., $c(q)$ in Equation \eqref{eq:context}, to the LambdaMart model. 

We first cluster the context vectors to $n$ clusters with k-means \footnote{Other clustering methods can be used as well. We leave it as future work. } and obtain a one-hot vector of length $n$, denoted as $cluster(q) \in \mathbb{R}^{n \times 1}$. The dimension of the corresponding  cluster id of the context has value 1 and the rest dimensions are all 0. To let the context take more effect, instead of adding $cluster(q)$ as features to a LambdaMart model directly, we bundle the document and q-d matching features with the $cluster(q)$. Specifically, we extract $l$ most important features in the feature set $[F_D(d);F_{QD}(q_0,d)]$, denoted as $F_l(q,d)\in \mathbb{R}^{l \times 1}$, do $F_l(q,d) \cdot cluster(q)^T$, flatten the yielded matrix to a vector $F_{fc}$ of length $l*n$, and add $F_{fc}$ to LambdaMart as additional features. 
In our experiments, $l$ is set to 3 and $F_l(q,d)$ includes recency, email length, and the overall BM25 score computed with multiple fields (BM25f) \cite{robertson2004simple}.

We also tried other methods of incorporating the context into LambdaMart, such as adding the context vector as $2m$ features directly, adding the cluster id as a single feature, and adding $n$ features either with a one-hot vector according to the cluster id or the probability of the context belonging to each cluster. These approaches are worse than the method we propose so we do not include the corresponding experimental results in the paper. 

\section{Experimental Setup}
\label{sec:exp_setup}
\subsection{Data \& Privacy Statement}
\label{subsec:dataset}
We use search logs obtained from one of the world's largest email search engines to collect the data. To protect user privacy, the raw text of their queries and documents cannot be accessed. Only features extracted from specifically designed privacy-preserved pipelines are available, as mentioned in Section \ref{sec:method}. Each candidate document has over 300 associated features. We randomly sampled some users from two-week search logs and only kept the users who have issued more than 10 queries during this period. We observe that over 10\% of users satisfy this condition. To reduce the effect of the uncommon users with too many queries, we filtered out 90\% of the users who issued more than 20 queries. In this way, we obtained hundreds of thousands of queries associated with the users in total. There are about 30 candidate documents on average and at most 50 documents for each query, out of which about 3 documents have positive labels on average. The labels are extracted based on users' interaction with the documents, such as click, reply, forward, long-dwell, etc., and there are five levels in total - bad, fair, good, excellent, and perfect. At each level, the average number of documents given a query is 26.91, 1.87, 0.03, 0.55, and 0.64, respectively.

Note that the dataset we produce has no user content, and it is also anonymized and delinked from the end-user. Under the End-user License Agreement (EULA), the dataset can be used to improve the service quality.

\textbf{Data Partition.}
To obtain data partitions consistent with real scenarios, we ordered the queries by their search time and divided the data to the training/validation/test set according to 0.8:0.1:0.1 in chronological order. Experiments on both neural models and LambdaMart models are based on this data. 

To see whether CNRM can generalize well to unseen users and achieve better search quality based on numerical ranking features in their search history, we also partition the data according to users. Specifically, we randomly split all the users to training/validation/test according to 0.8:0.1:0.1 and put their associated queries to the corresponding partitions. The users in each partition do not overlap. 

\subsection{Evaluation}
\label{subsec:eval}
We conduct two types of evaluation to see whether users' historical behaviors benefit search quality. Specifically, we compare NRM to CNRM and the vanilla LambdaMart model to the LambdaMart model with context added as additional features (as stated in Section \ref{subsec:context2lambdamart}). These comparisons are based on the data partitioned in chronological order, which is consistent with real scenarios. 
Also, we compare CNRM with NRM on data partitioned according to users to show the ability of CNRM generalizing to unseen users. 
When we conduct comparisons between neural models with and without incorporating the context, we also include several other baselines:
\begin{itemize}
    \item \textbf{BM25f}: BM25f measures the relevance of an email. It computes the overall BM25 score of an email based on word matching in its multiple fields \cite{craswell2005microsoft, robertson2004simple}.
    \item \textbf{Recency}: Recency ranks emails according to the time they were created, which is the ranking criterion for the recency panel of most email search engines.
    \item \textbf{CLSTM}: The Context-aware Long Short Term Memory model (CLSTM) encodes the context of user behavior features with LSTM \cite{hochreiter1997long} instead of Transformers \cite{vaswani2017attention}. In other words, CLSTM replaces the Transformers in CNRM with LSTM. 
\end{itemize}

Since top results matter more when users examine the emails and our data has multi-grade labels, we use Normalized Discounted Cumulative Gain (NDCG) at cutoff 3,5,10 as evaluation metrics. Two-tailed student's t-tests with p<0.05 are conducted to check statistical significance. 

\subsection{Training Settings}
\label{subsec:train_setting}
We implemented the neural models with PyTorch \footnote{\url{https://pytorch.org/}} and trained each neural model on a single NVIDIA Tesla K80 GPU. All the neural models were trained with 128 as batch size and for 10 epochs, within which they can converge well. We set embedding size of discrete features $e=10$ (in Equation \eqref{eq:qdiscrete1}) and the hidden dimension $m$ in Section \ref{sec:method} to 64. The performances do not improve with larger embedding sizes so we only report results with $m=64$. For the Transformers in CNRM, we sweep the number of attention heads from \{1,2,4,8\}, the transformer layers from 1 to 3, and the dimension size of the feed-forward sub-layer from \{128, 256, 512\}. For the CLSTM model, we tried layer numbers from 1 to 3. We set the history length $k$ to 10. We use Adam with $\beta_1=0.9, \beta_2=0.999$ to optimize the neural models. The learning rate is initially set to 0.002 and then warmed up over the first \{2,000, 3,000, 4,000\} steps, following the paradigm in \cite{vaswani2017attention}. The coefficient for regularization Equation \eqref{eq:loss} is selected from \{1e-5, 5e-5, 1e-6\}.

We train the LambdaMart models using LightGBM \footnote{\url{https://lightgbm.readthedocs.io/en/latest/}}. We set the number of trees to be 500, the number of leaves in each tree as 150, the shrinkage rate as 0.3, the objective as LambdaRank, and the rounds of early-stop based on the validation performance as 30. We use default parameters for the other settings. The number of clusters $n$ is set to 10. 
\section{Results and Discussion}
\label{sec:results}
In this section, we first show the experimental results of neural models, conduct ablation studies, and analyze what information the context vector captures. Then we show the results of LambdaMart models with the context information added and give some analysis accordingly. 

\subsection{Neural Model Evaluation}
\label{subsec:neural_eval}
\textbf{Overall Performance.} 
Table \ref{tab:neural_results} shows that CNRM and CLSTM outperform NRM on both versions of data, i.e., partitioned according to time or users. This indicates that encoding the ranking features corresponding to users' historical queries and positive documents can benefit the neural ranking model. This information improves search quality not only for the future queries of the same user but also for an unseen user's queries given his/her previous queries. CNRM has better performances than CLSTM, showing the superiority of Transformers over the recurrent neural networks, similar to the observation in \cite{vaswani2017attention}. 

BM25f and Recency both have significantly degraded performances compared to NRM. It is not surprising since they are used together with other features in the neural models. Also, Recency outperforms BM25f by a large margin. In contrast to Web search where relevance usually matters the most, recency plays a more crucial role than relevance in email search. This observation is consistent with the fact that recency has been the criterion for ranking emails in traditional email search engines and the recency panel in most current email search engines. 

NRM has worse NDCG@3,5,10 but better NDCG@1 than the vanilla LambdaMart model on both versions of data. We have not included the performances of vanilla LambdaMart in Table \ref{tab:neural_results} since Table \ref{tab:neural_results} is for comparison between heuristic methods and neural models. 
NRM has 6.65\% better NDCG@1, while 7.99\%, 8.90\%, and 7.78\% worse NDCG@3,5, and 10, than LambdaMart on the data partitioned according to time. On the data partitioned according to users, NRM has 7.40\% better NDCG@1 but 5.02\%, 5.42\%, and 4.53\% worse NDCG@3,5,10 than vanilla LambdaMart. 
With the same feature set, the advantages of LambdaMart, such as the gradient boosting mechanism, the NDCG-aware loss function, and the ability to handle continuous features, make it hard to beat by a neural model.


\begin{table}
	\caption{Comparisons between CNRM and the baselines on the data partitioned according to time and users. The reported numbers are the improvements over NRM. `*' and `$\dag$' indicate statistically significant improvements over NRM and CLSTM respectively. }
	\centering
	\label{tab:neural_results}
	\begin{tabular}{l| l l l l}
		\hline
        Partition & Method & NDCG@3 & NDCG@5 & NDCG@10 \\ 
        \hline
        \multirow{5}{*}{Time}
        & BM25f & -51.55\% & -46.95\% & -41.37\% \\
        & Recency & -38.32\% & -32.27\% & -26.69\% \\
        & NRM & +0.00\% & +0.00\% & +0.00\% \\
        \cline{2-5}
        & CLSTM & +1.29\%* & +1.35\%* & +1.51\%* \\
        & CNRM & \textbf{+2.52\%$^{*\dag}$} & \textbf{+2.89\%$^{*\dag}$} & \textbf{+2.88\%$^{*\dag}$} \\
        \hline
        \multirow{5}{*}{Users}
        & BM25f & -51.17\% & -46.61\% & -41.37\% \\
        & Recency & -37.23\% & -31.58\% & -25.94\% \\
        & NRM & +0.00\% & +0.00\% & +0.00\% \\
        \cline{2-5}
        & CLSTM & +1.79\%* & +2.15\%* & +2.28\%* \\
        & CNRM & \textbf{+2.22\%$^{*\dag}$} & \textbf{+2.60\%$^{*\dag}$} & \textbf{+2.75\%$^{*\dag}$} \\
        \hline
	\end{tabular}
\end{table}

\begin{table}
	\caption{Performance improvements of the variants of CNRM over NRM. $F_Q$, $F_D$, and $F_{QD}$ indicates the CNRM with only query, document, and q-d matching features encoded. * indicates statistically significant differences with NRM.}
	\centering
	\large
	\label{tab:ablation_results}
	\begin{tabular}{l l l l}
	\hline
    Model & NDCG@3 & NDCG@5 & NDCG@10 \\
    \hline
    Full CNRM & \textbf{+2.52\%*} & \textbf{+2.89\%*} & \textbf{+2.88\%*} \\ 
    \hline
    with $F_{Q}$ & +1.12\%* & +1.19\%* & +1.11\%* \\ 
    with $F_{D}$ & +1.71\%* & +2.03\%* & +2.17\%* \\ 
    with $F_{QD}$ & +1.95\%* & +2.25\%* & +2.25\%* \\ 
    with $F_{D}\&F_{QD}$ & +2.14\%* & +2.49\%* & +2.56\%* \\ 
    \hline
    w/o posEmb & +0.72\%* & +0.99\%* & +1.15\%* \\ 
    \hline
	\end{tabular}
\end{table}

\begin{figure*}
	\centering
	\begin{subfigure}{.48 \textwidth}
		\includegraphics[width=3.7in]{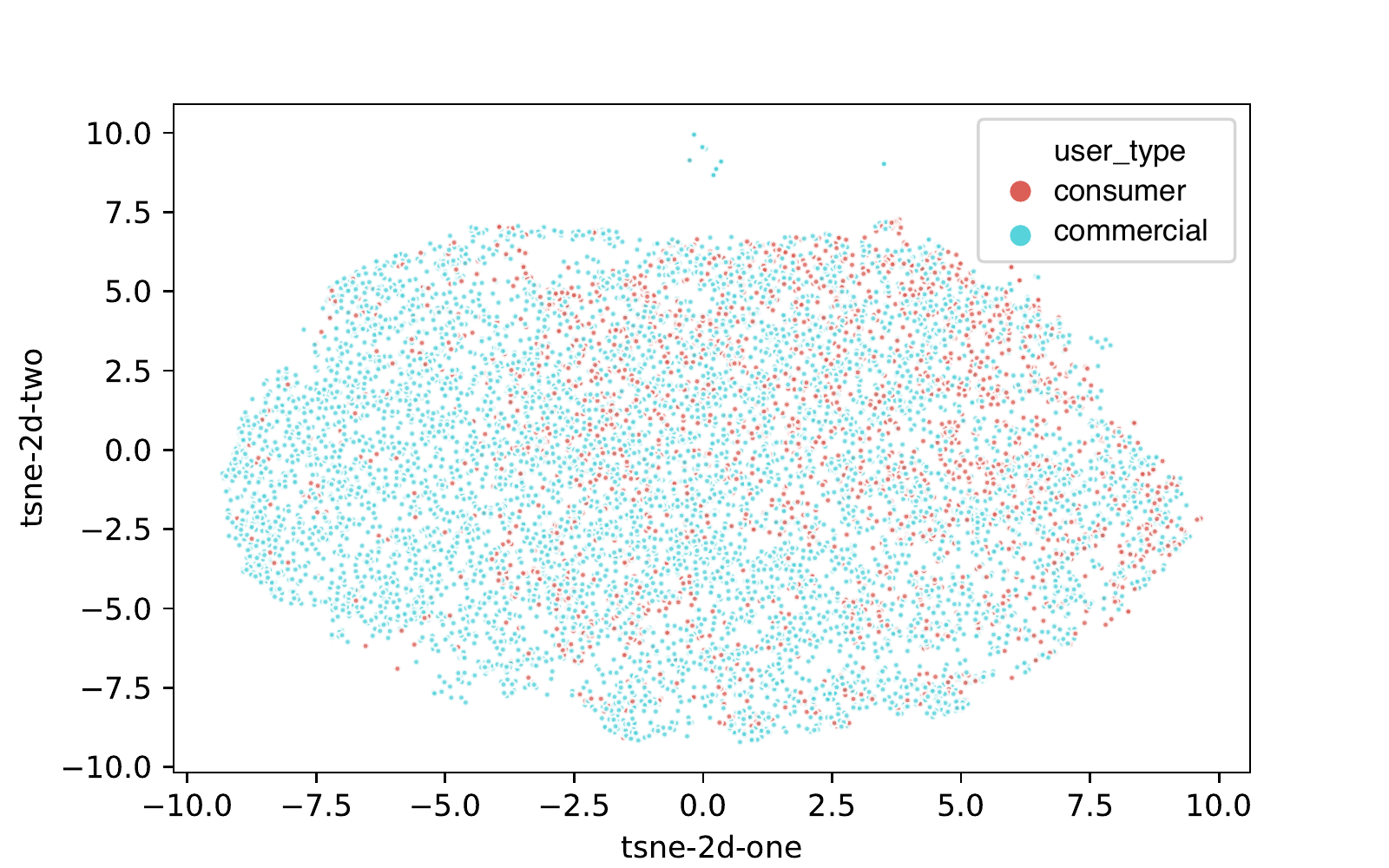}
		\caption{Context vectors encoded with $F_{QD}$ w.r.t. user type}
		\label{fig:qd1}
	\end{subfigure}%
	\begin{subfigure}{.48 \textwidth}
		\includegraphics[width=3.7in]{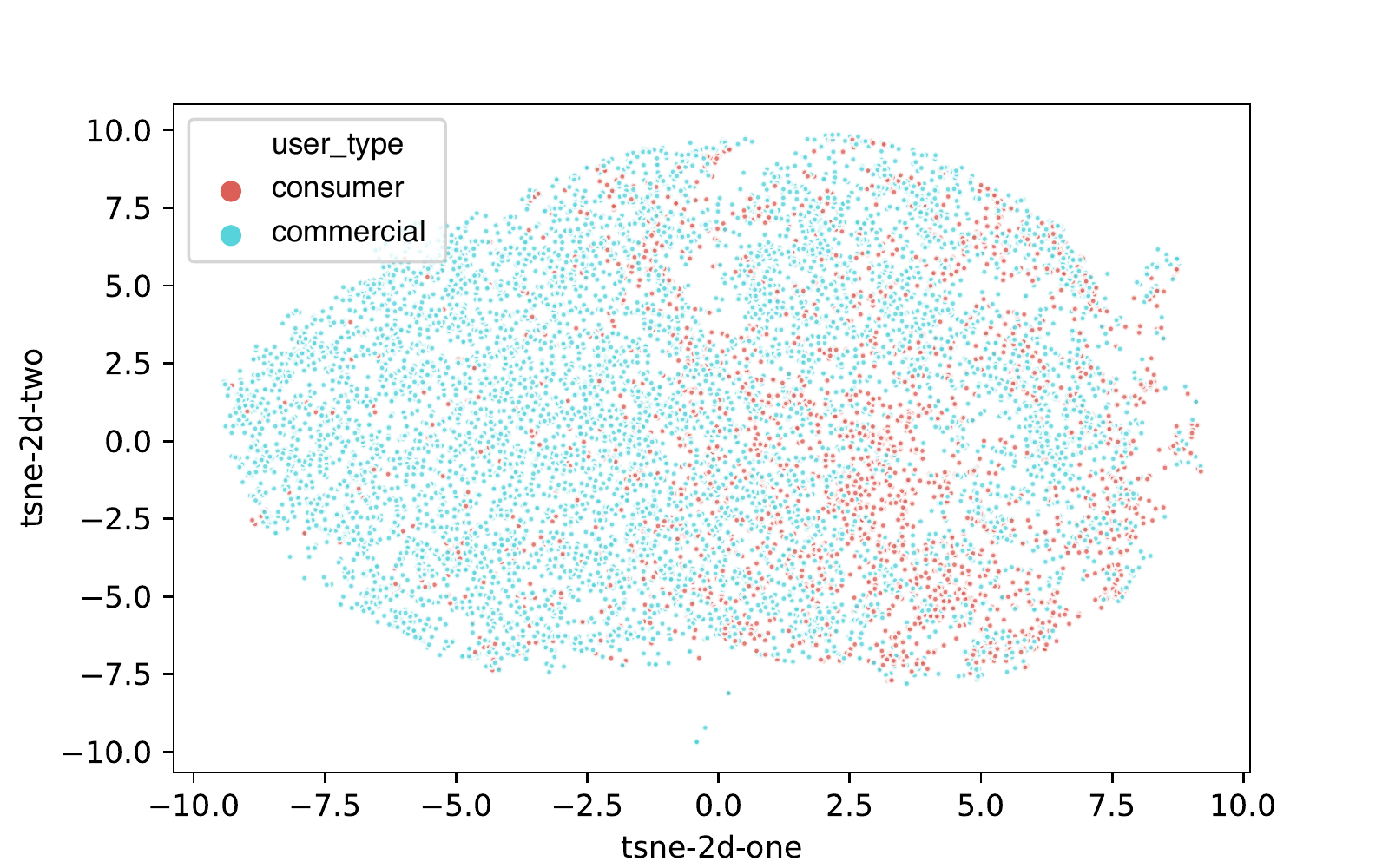}
		\caption{Context vectors encoded with $F_D$ and $F_{QD}$ w.r.t. user type}
		\label{fig:d1qd1}
	\end{subfigure}%
	\hfill
	\begin{subfigure}{.48 \textwidth}
		\includegraphics[width=3.7in]{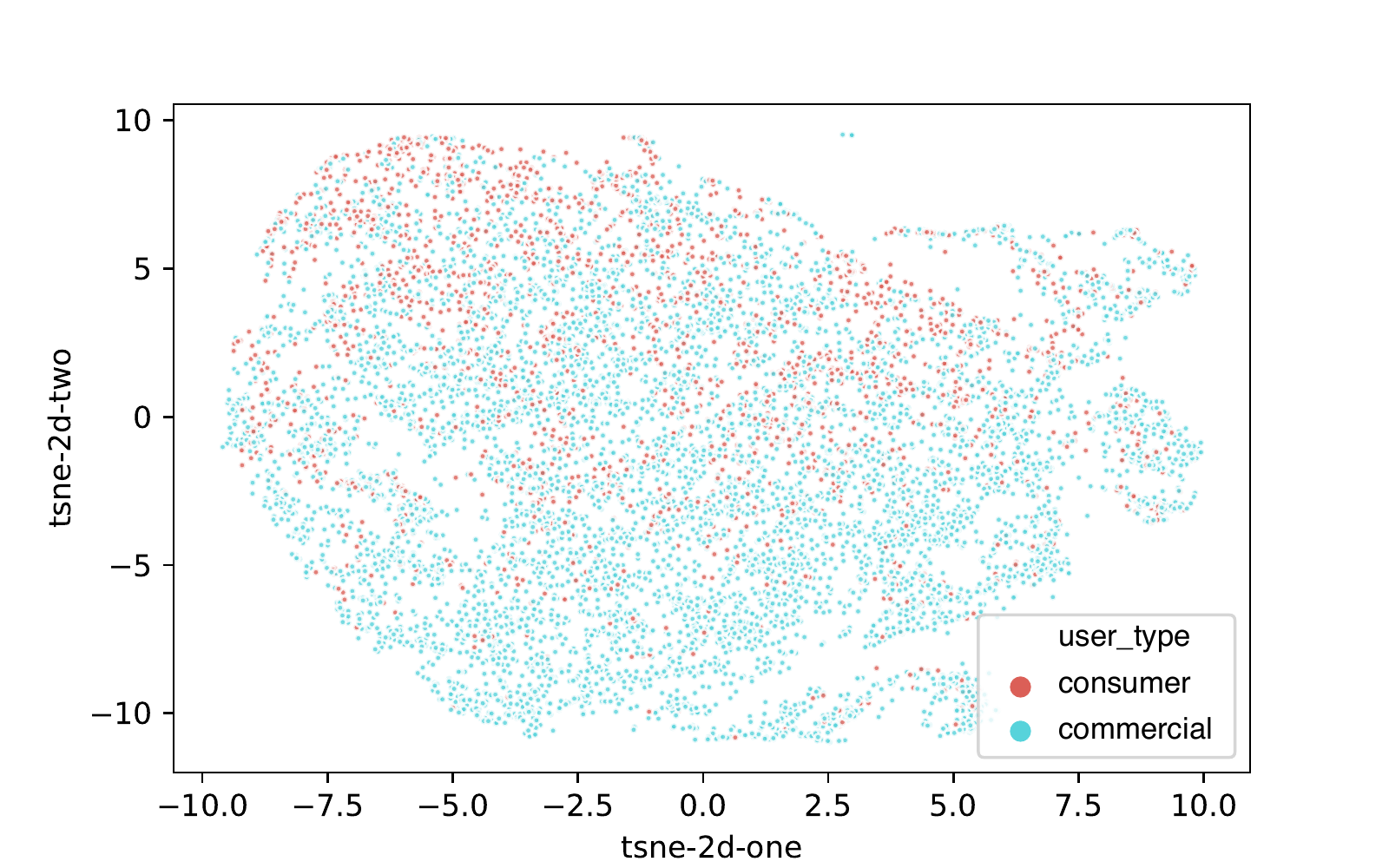}
		\caption{Context vectors encoded with $F_D$ w.r.t. user type}
		\label{fig:d1}
	\end{subfigure}%
	\begin{subfigure}{.48 \textwidth}
		\includegraphics[width=3.7in]{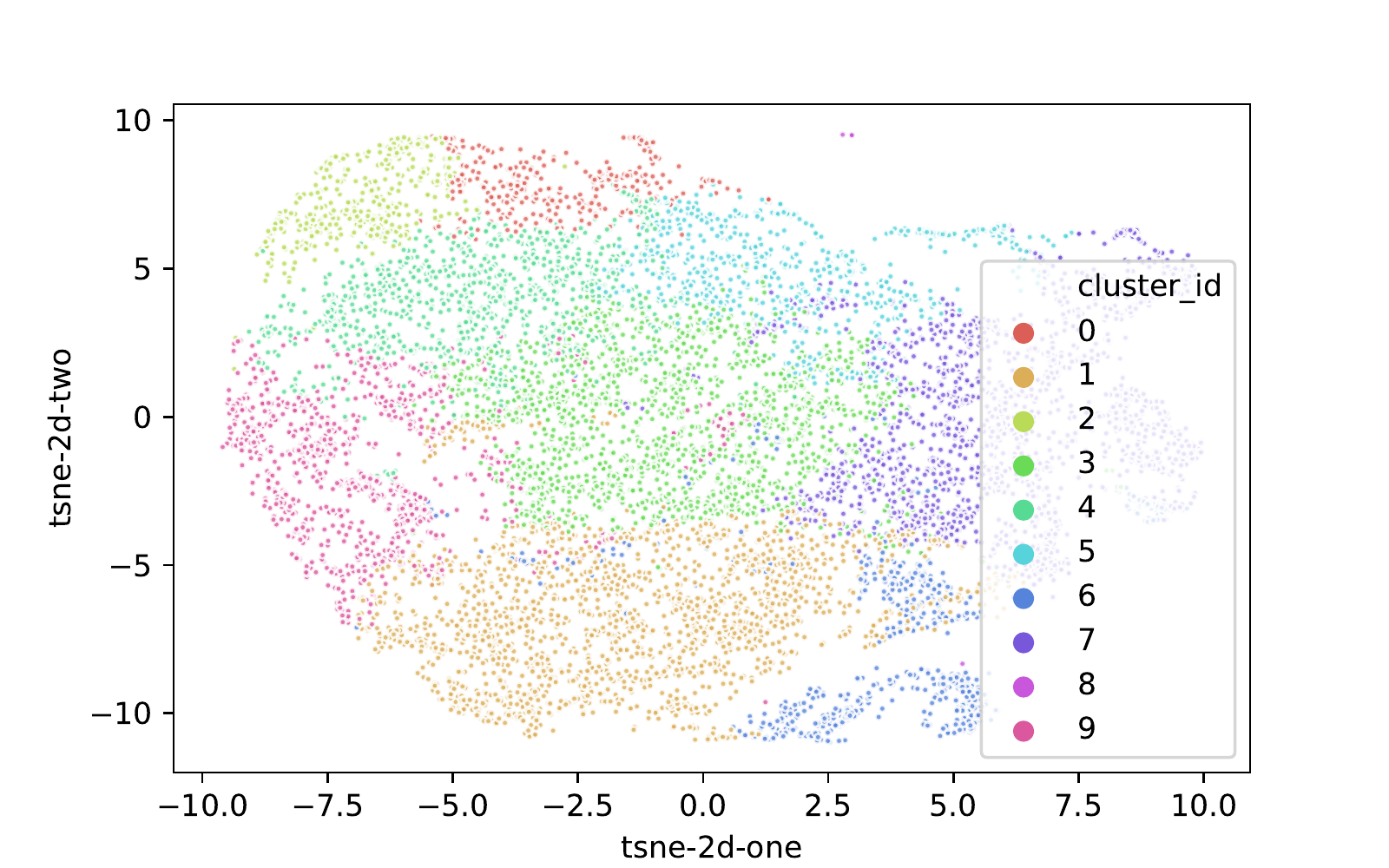}
		\caption{Context vectors encoded with $F_D$ w.r.t. cluster ID}
		\label{fig:d1_clusters}
	\end{subfigure}%
	\caption{2-D visualization of context vectors in CNRM by t-SNE \cite{Maaten08visualizingdata} with respect to user\_type and cluster ID.} 
	\label{fig:context_vec}
\end{figure*}
\textbf{Ablation Study.}
We conduct ablation studies on the data partitioned according to time by keeping only some groups of features and remove the others when encoding the context to study their contributions. In other words, we only use the query features $F_Q$, the document features $F_D$, the q-d matching features $F_{QD}$, or both $F_D$ and $F_{QD}$ corresponding to the historical queries and associated positive documents during context encoding. The current query features are also excluded during encoding (See Figure \ref{fig:model}) so that the differences between CNRM and NRM only come from the encoding of certain groups of features.
As shown in Table \ref{tab:ablation_results}, each group of feature alone has a contribution, among which q-d matching features contribute the most and query features contribute the least. It is consistent with our intuition since query features indicate the property of the query or the user while document and q-d matching features carry the most critical information to determine whether a document is a target or not, such as recency and relevance. Incorporating both $F_D$ and $F_{QD}$ leads to better performances while still worse than full CNRM with all the components. 

We also remove the position embeddings before feeding the sequence to transformer layers to see whether the ordering of the sequence matters. Results show that the performance will drop significantly without position embeddings, indicating that the ordering of users' historical behaviors is important during encoding. 

In addition, without unbiased learning, for both NRM and CNRM, the performance will degrade by about 15\% of the models using unbiased learning. The improvement percentage of CNRM over NRM is similar to the numbers shown in Table \ref{tab:ablation_results}. 

\textbf{Context Vector Analysis.}
To see whether the context vector $c(q_0)$ (in Figure \ref{fig:model}) captures some information that can differentiate users, we try to visualize $c(q_0)$ of each query in terms of a user type. As we mentioned in Section \ref{sec:introduction}, consumer and commercial users have significantly different behavioral patterns. We want to check whether the differences can be revealed when CNRM only encodes document features ($F_D$), q-d matching features ($F_{QD})$, or both of them. Meanwhile, the query-features of the current query $q_0$ in Figure \ref{fig:model} is replaced by a trainable vector of dimension $2m$. This ensures that all query features that include the user type (consumer or commercial) are excluded during encoding and $c(q_0)$ has not seen the user type in advance. 

Based on the data partitioned according to time, we use t-SNE \cite{Maaten08visualizingdata}, which can reserve both the local and global structures of the high-dimensional data well when mapping to low-dimensional space as discussed in \cite{Maaten08visualizingdata}, to map $c(q_0)$ to the 2-d plane. We randomly sample 10,000 queries and obtain their corresponding context vectors with the same CNRMs that are evaluated in Table \ref{tab:ablation_results}. Figure \ref{fig:qd1}, \ref{fig:d1qd1}, and \ref{fig:d1} shows the 2-d visualization of context vectors encoded with $F_{QD}$, both $F_{QD}$ and $F_D$, and $F_D$ in terms of user type respectively. 

We observe that by encoding $F_{QD}$/$F_D$ alone, $c(q_0)$ can differentiate commercial and consumer users to some extent by putting most of consumer users to the right/upper-left part of Figure \ref{fig:qd1}/\ref{fig:d1}. With both $F_{D}$ and $F_{QD}$ encoded, $c(q_0)$ further pushes consumer users to the right part of Figure \ref{fig:d1qd1}. These observations indicate that the context vectors of CNRM can differentiate consumer and commercial users by encoding the document and q-d matching features of their positive historical documents. The two groups of features take a decent effect individually to capture their different behavioral patterns and are more discriminative when used together. It is worth mentioning that the context vectors have captured information beyond whether the user is a commercial or consumer user. This can be verified by the fact that the user type of the current query ($q_0$) is already used in the candidate document for ranking and the learned context still leads to significantly better performances, as shown in Section \ref{subsec:neural_eval}. 

\subsection{LambdaMart Model Evaluation}
In common practice, LambdaMart has been widely used to aggregate multiple features to serve online query requests due to its state-of-the-art performance and high efficiency. So it is very important to find out whether the context information learned in CNRM can benefit the LambdarMart model.  
The baseline LambdaMart model was trained with the same feature set mentioned in Section \ref{sec:method}. For any version of CNRM to be evaluated, the context vector for each query was extracted and put to one of the 10 clusters; the top 3 features (Recency, EmailLength, and BM25f) are bundled with each cluster and added to LambdaMart as additional features, as stated in Section \ref{subsec:context2lambdamart}. Then a new LambdarMart model is trained and evaluated against the baseline version.

\textbf{Overall Performances.}
Based on the data partitioned according to time (See Section \ref{subsec:dataset}), the context information from four versions of CNRM is introduced to the LambdarMart model, which are CNRM that only encodes document features ($F_D$), q-d matching features ($F_{QD}$), both $F_D$ and $F_{QD}$, and full information in the context, same as the settings in our ablation study in Section \ref{subsec:neural_eval}. We did not include the context encoded with query features $F_Q$ alone since user-level information such as user type is the same for all user queries and contains less extra information than $F_D$ and $F_{QD}$. 

Since the clustering process forces the representations to degrade from dense to discrete, much information could be lost. In addition, query-level features usually take less effect than document-level or q-d matching features since they are the same for all the candidate documents under a query. Thus, it is challenging to make the query context work effectively in the LambdaMart model. 
Nevertheless, as shown in Table \ref{tab:lambdamart}, the context learned from CNRM that encodes $F_D$ alone leads to significantly better performances than the baseline LambdaMart model, which is also the best performances among all. 

From Table \ref{tab:lambdamart}, we observe that the improvements each method obtains in the LambdaMart model do not follow the same trends of their improvements over the baseline NRM shown in Table \ref{tab:ablation_results}. This is not surprising since the clustering process could lose some information and the condensed query cluster information may have uncertain overlap with existing user properties such as user type and mailbox locale, which already take effect in LambdaMart. 

\begin{table}
	\caption{Performance improvements of the LambdaMart model with the context cluster information over the baseline version without context on the data partitioned according to time. * indicates statistically significant differences.}
	\centering
	\label{tab:lambdamart}
	\begin{tabular}{l l l l}
	\hline
    Model & NDCG@3 & NDCG@5 & NDCG@10 \\
    \hline
    + Context with $F_{D}$ & +0.60\%* & +0.49\%* & +0.56\%* \\
    + Context with $F_{QD}$ & +0.34\% & +0.35\% & +0.28\% \\
    + Context with $F_{D}\&F_{QD}$ & +0.33\% & +0.41\% & +0.30\% \\
    + Context from Full CNRM & +0.37\% & +0.20\% & +0.18\% \\
    \hline
	\end{tabular}
\end{table}

\begin{figure}
	\centering
    \includegraphics[width=0.48\textwidth]{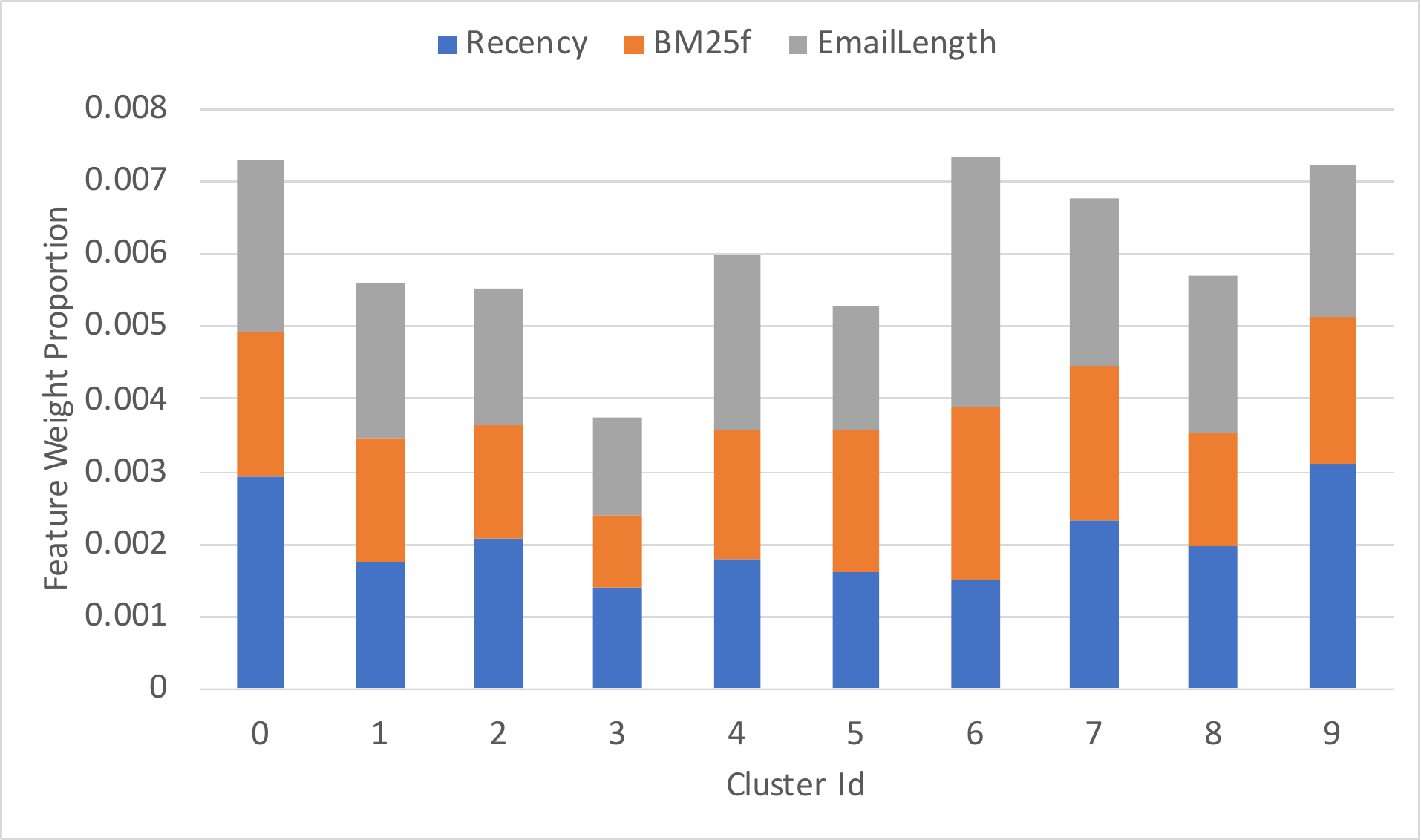} 
	\caption{Feature weights of the added features (Recency, EmailLength, and BM25f) in the LambdaMart model that uses context encoded with $F_D$ in terms of cluster ID. }
	\label{fig:feature_weight}
\end{figure}

\textbf{Feature Weight Analysis}.
In the LambdaMart models, the top 3 features with the most feature weight are Recency, BM25f, and EmailLength, which occupy about 24\%, 5\%, and 4\% of the total feature weights. These features could play a different role in terms of different query clusters. To compare the weights of these 3 features when bundled with each cluster, we plot the feature weights of the LambdaMart model that has significant improvements over the baseline, which is incorporating the context encoded with $F_{D}$ alone, in Figure \ref{fig:feature_weight}. Cluster 6 has the most overall feature weights and has a significantly different feature weight distribution than the other clusters. Queries in Cluster 6 emphasize email length the most and recency the least, which indicates that this cluster requires different matching patterns from the global distribution. As shown in Figure \ref{fig:d1_clusters}, Cluster 6 is located in the lower-right corner of the figure, which has a little gap with the other points. This may also indicate that Cluster 6 is quite different from the other clusters. 

\begin{figure}
	\centering
    \includegraphics[width=0.45\textwidth]{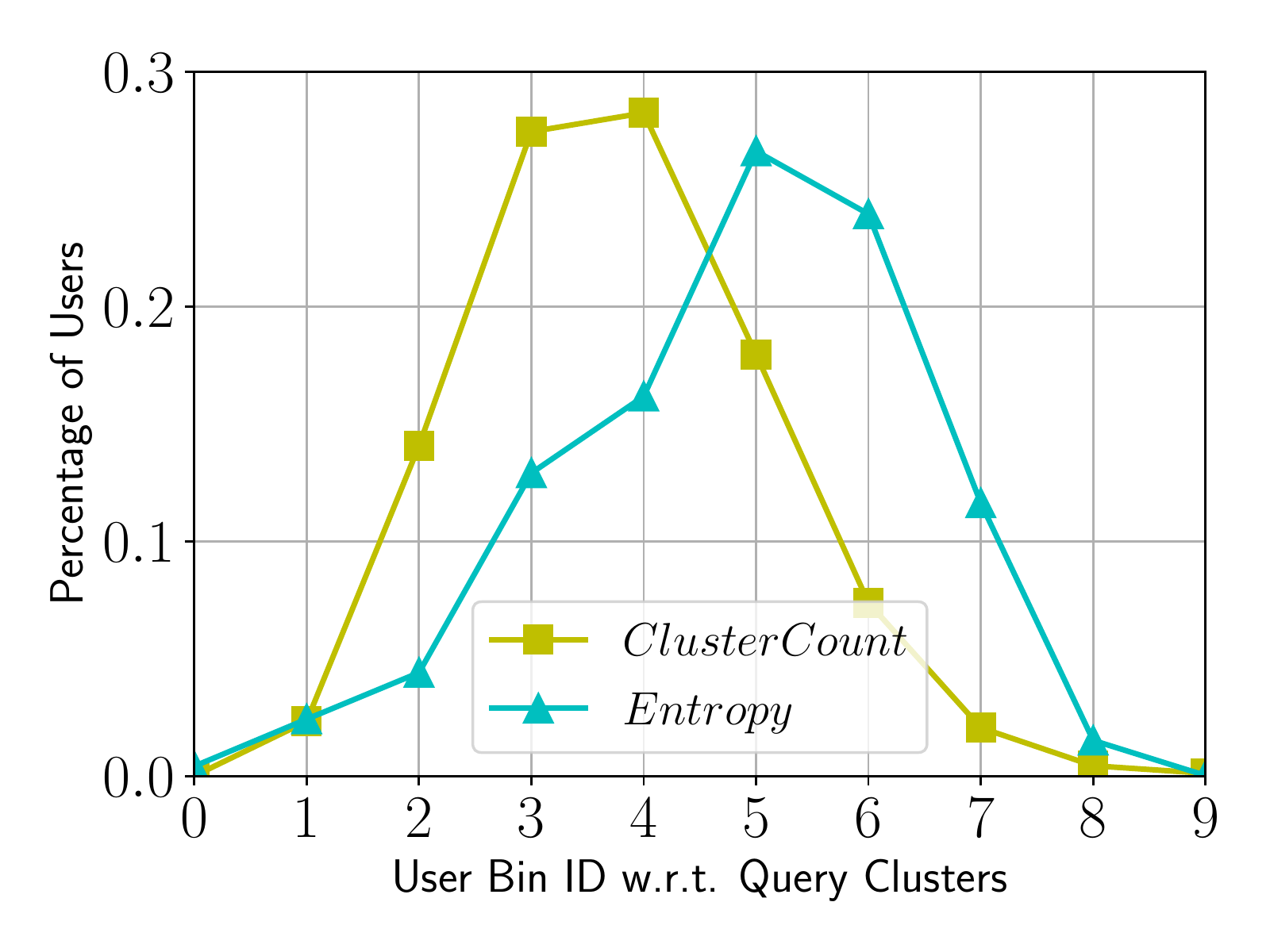} 
	\caption{The distribution of users when we put each user to 10 bins according to the count of unique clusters the user's queries belongs to and the entropy of the user's query cluster distribution.}
	\label{fig:uqcluster_distr}
\end{figure}

\textbf{User Distribution w.r.t. Their Query Clusters.}
In this part, we study how users are distributed in terms of the clusters their queries belong to. We aim to answer the following question: do the query contexts associated with the same user concentrate or scatter in the latent space? We use two criteria to analyze the corresponding user distribution: (1) the number of unique clusters a user's queries belong to, (2) the entropy of the cluster distribution of queries associated with a user. The entropy criterion differentiates query counts in each cluster while the cluster count criterion does not. 
Then we divide the possible values of cluster count and entropy into 10 bins and put users into each bin to see the distribution. Figure \ref{fig:uqcluster_distr} shows the user distribution in percentage regarding the two criteria using the CNRM that only encodes $F_D$ in the context. 

We have not observed any pronounced patterns from the user distribution. The two resulting distributions are similar. Most users have 4 or 5 query clusters (Bin 3 or 4) and fall into Bin 5 or 6 in terms of entropy. Most users do not have concentrated query clusters, indicating that user queries could have dynamic context along with time. This makes sense since the context learned in CNRM is related to not only the users' static properties but also their issued queries that could be diversified. We leave the study of extracting users' static properties and the effect of the user cluster information as our future work.

\section{Conclusion and Future Work}
\label{sec:conclusion}
This paper presents the first comprehensive work to improve the email search quality based on ranking features in the user search history. We investigate this problem on both neural models and the state-of-the-art learning-to-rank model - LambdaMart \cite{burges2010ranknet}. 
We leverage the ranking features of users' historical queries and the corresponding positive emails to characterize the users and propose a context-dependent neural ranking model (CNRM) that encodes the sequence of the ranking features with a transformer architecture. For the LambdaMart model, we cluster the query context vectors obtained from CNRM and incorporate the cluster information into LambdaMart. Based on the dataset constructed from the search log of one of the world's largest email search engines, we show that for both neural models and the LambdaMart model, incorporating this query context could improve the search quality significantly.

As a next step, we would like to investigate using another optimization goal in addition to the current ranking optimization to enhance the neural ranker, which is differentiating the query from the same user and the other users given the query context encoded from a user's search history. We are also interested in studying how to incorporate the query context into LambdaMart more effectively, such as investigating other clustering methods including neural end-to-end clustering approaches and incorporating the context information differently. As mentioned earlier, extracting users' static properties and use them as query context is another interesting direction.

\begin{acks}
This work was supported in part by the Center for Intelligent Information Retrieval. Any opinions, findings and conclusions or recommendations expressed in this material are those of the authors and do not necessarily reflect those of the sponsor.
\end{acks}

\bibliographystyle{ACM-Reference-Format}
\balance
\bibliography{reference}


\begin{thebibliography}{38}


\ifx \showCODEN    \undefined \def \showCODEN     #1{\unskip}     \fi
\ifx \showDOI      \undefined \def \showDOI       #1{#1}\fi
\ifx \showISBNx    \undefined \def \showISBNx     #1{\unskip}     \fi
\ifx \showISBNxiii \undefined \def \showISBNxiii  #1{\unskip}     \fi
\ifx \showISSN     \undefined \def \showISSN      #1{\unskip}     \fi
\ifx \showLCCN     \undefined \def \showLCCN      #1{\unskip}     \fi
\ifx \shownote     \undefined \def \shownote      #1{#1}          \fi
\ifx \showarticletitle \undefined \def \showarticletitle #1{#1}   \fi
\ifx \showURL      \undefined \def \showURL       {\relax}        \fi
\providecommand\bibfield[2]{#2}
\providecommand\bibinfo[2]{#2}
\providecommand\natexlab[1]{#1}
\providecommand\showeprint[2][]{arXiv:#2}

\bibitem[\protect\citeauthoryear{AbdelRahman, Hassan, and Bahgat}{AbdelRahman
  et~al\mbox{.}}{2010}]%
        {abdelrahman2010new}
\bibfield{author}{\bibinfo{person}{Samir AbdelRahman}, \bibinfo{person}{Basma
  Hassan}, {and} \bibinfo{person}{Reem Bahgat}.}
  \bibinfo{year}{2010}\natexlab{}.
\newblock \showarticletitle{A new email retrieval ranking approach}.
\newblock \bibinfo{journal}{\emph{arXiv preprint arXiv:1011.0502}}
  (\bibinfo{year}{2010}).
\newblock


\bibitem[\protect\citeauthoryear{Aberdeen, Pacovsky, and Slater}{Aberdeen
  et~al\mbox{.}}{2010}]%
        {aberdeen2010learning}
\bibfield{author}{\bibinfo{person}{Douglas Aberdeen}, \bibinfo{person}{Ondrey
  Pacovsky}, {and} \bibinfo{person}{Andrew Slater}.}
  \bibinfo{year}{2010}\natexlab{}.
\newblock \showarticletitle{The learning behind gmail priority inbox}.
\newblock  (\bibinfo{year}{2010}).
\newblock


\bibitem[\protect\citeauthoryear{Ai, Bi, Guo, and Croft}{Ai
  et~al\mbox{.}}{2018a}]%
        {ai2018learning}
\bibfield{author}{\bibinfo{person}{Qingyao Ai}, \bibinfo{person}{Keping Bi},
  \bibinfo{person}{Jiafeng Guo}, {and} \bibinfo{person}{W~Bruce Croft}.}
  \bibinfo{year}{2018}\natexlab{a}.
\newblock \showarticletitle{Learning a Deep Listwise Context Model for Ranking
  Refinement}.
\newblock \bibinfo{journal}{\emph{arXiv preprint arXiv:1804.05936}}
  (\bibinfo{year}{2018}), \bibinfo{pages}{135--144}.
\newblock


\bibitem[\protect\citeauthoryear{Ai, Bi, Luo, Guo, and Croft}{Ai
  et~al\mbox{.}}{2018b}]%
        {ai2018unbiased}
\bibfield{author}{\bibinfo{person}{Qingyao Ai}, \bibinfo{person}{Keping Bi},
  \bibinfo{person}{Cheng Luo}, \bibinfo{person}{Jiafeng Guo}, {and}
  \bibinfo{person}{W~Bruce Croft}.} \bibinfo{year}{2018}\natexlab{b}.
\newblock \showarticletitle{Unbiased Learning to Rank with Unbiased Propensity
  Estimation}.
\newblock \bibinfo{journal}{\emph{arXiv preprint arXiv:1804.05938}}
  (\bibinfo{year}{2018}).
\newblock


\bibitem[\protect\citeauthoryear{Bendersky, Wang, Metzler, and
  Najork}{Bendersky et~al\mbox{.}}{2017}]%
        {bendersky2017learning}
\bibfield{author}{\bibinfo{person}{Michael Bendersky}, \bibinfo{person}{Xuanhui
  Wang}, \bibinfo{person}{Donald Metzler}, {and} \bibinfo{person}{Marc
  Najork}.} \bibinfo{year}{2017}\natexlab{}.
\newblock \showarticletitle{Learning from user interactions in personal search
  via attribute parameterization}. In \bibinfo{booktitle}{\emph{Proceedings of
  the Tenth ACM International Conference on Web Search and Data Mining}}.
  \bibinfo{pages}{791--799}.
\newblock


\bibitem[\protect\citeauthoryear{Bennett, White, Chu, Dumais, Bailey, Borisyuk,
  and Cui}{Bennett et~al\mbox{.}}{2012}]%
        {bennett2012modeling}
\bibfield{author}{\bibinfo{person}{Paul~N Bennett}, \bibinfo{person}{Ryen~W
  White}, \bibinfo{person}{Wei Chu}, \bibinfo{person}{Susan~T Dumais},
  \bibinfo{person}{Peter Bailey}, \bibinfo{person}{Fedor Borisyuk}, {and}
  \bibinfo{person}{Xiaoyuan Cui}.} \bibinfo{year}{2012}\natexlab{}.
\newblock \showarticletitle{Modeling the impact of short-and long-term behavior
  on search personalization}. In \bibinfo{booktitle}{\emph{Proceedings of the
  35th international ACM SIGIR conference on Research and development in
  information retrieval}}. \bibinfo{pages}{185--194}.
\newblock


\bibitem[\protect\citeauthoryear{Bhole and Udupa}{Bhole and Udupa}{2015}]%
        {bhole2015correcting}
\bibfield{author}{\bibinfo{person}{Abhijit Bhole} {and}
  \bibinfo{person}{Raghavendra Udupa}.} \bibinfo{year}{2015}\natexlab{}.
\newblock \showarticletitle{On correcting misspelled queries in email search}.
  In \bibinfo{booktitle}{\emph{Proceedings of the Twenty-Ninth AAAI Conference
  on Artificial Intelligence}}. \bibinfo{pages}{4266--4267}.
\newblock


\bibitem[\protect\citeauthoryear{Burges}{Burges}{2010}]%
        {burges2010ranknet}
\bibfield{author}{\bibinfo{person}{Christopher~JC Burges}.}
  \bibinfo{year}{2010}\natexlab{}.
\newblock \showarticletitle{From ranknet to lambdarank to lambdamart: An
  overview}.
\newblock \bibinfo{journal}{\emph{Learning}} \bibinfo{volume}{11},
  \bibinfo{number}{23-581} (\bibinfo{year}{2010}), \bibinfo{pages}{81}.
\newblock


\bibitem[\protect\citeauthoryear{Carmel, Halawi, Lewin-Eytan, Maarek, and
  Raviv}{Carmel et~al\mbox{.}}{2015}]%
        {carmel2015rank}
\bibfield{author}{\bibinfo{person}{David Carmel}, \bibinfo{person}{Guy Halawi},
  \bibinfo{person}{Liane Lewin-Eytan}, \bibinfo{person}{Yoelle Maarek}, {and}
  \bibinfo{person}{Ariel Raviv}.} \bibinfo{year}{2015}\natexlab{}.
\newblock \showarticletitle{Rank by time or by relevance? Revisiting email
  search}. In \bibinfo{booktitle}{\emph{Proceedings of the 24th ACM
  International on Conference on Information and Knowledge Management}}.
  \bibinfo{pages}{283--292}.
\newblock


\bibitem[\protect\citeauthoryear{Carmel, Lewin-Eytan, Libov, Maarek, and
  Raviv}{Carmel et~al\mbox{.}}{2017}]%
        {carmel2017promoting}
\bibfield{author}{\bibinfo{person}{David Carmel}, \bibinfo{person}{Liane
  Lewin-Eytan}, \bibinfo{person}{Alex Libov}, \bibinfo{person}{Yoelle Maarek},
  {and} \bibinfo{person}{Ariel Raviv}.} \bibinfo{year}{2017}\natexlab{}.
\newblock \showarticletitle{Promoting relevant results in time-ranked mail
  search}. In \bibinfo{booktitle}{\emph{Proceedings of the 26th International
  Conference on World Wide Web}}. \bibinfo{pages}{1551--1559}.
\newblock


\bibitem[\protect\citeauthoryear{Craswell, De~Vries, and Soboroff}{Craswell
  et~al\mbox{.}}{2005a}]%
        {craswell2005overview}
\bibfield{author}{\bibinfo{person}{Nick Craswell}, \bibinfo{person}{Arjen~P
  De~Vries}, {and} \bibinfo{person}{Ian Soboroff}.}
  \bibinfo{year}{2005}\natexlab{a}.
\newblock \showarticletitle{Overview of the TREC 2005 Enterprise Track.}. In
  \bibinfo{booktitle}{\emph{Trec}}, Vol.~\bibinfo{volume}{5}.
  \bibinfo{pages}{1--7}.
\newblock


\bibitem[\protect\citeauthoryear{Craswell, Zaragoza, and Robertson}{Craswell
  et~al\mbox{.}}{2005b}]%
        {craswell2005microsoft}
\bibfield{author}{\bibinfo{person}{Nick Craswell}, \bibinfo{person}{Hugo
  Zaragoza}, {and} \bibinfo{person}{Stephen Robertson}.}
  \bibinfo{year}{2005}\natexlab{b}.
\newblock \showarticletitle{Microsoft cambridge at trec-14: Enterprise track}.
\newblock  (\bibinfo{year}{2005}).
\newblock


\bibitem[\protect\citeauthoryear{Croft and Wei}{Croft and Wei}{2005}]%
        {croft2005context}
\bibfield{author}{\bibinfo{person}{W~Bruce Croft} {and} \bibinfo{person}{Xing
  Wei}.} \bibinfo{year}{2005}\natexlab{}.
\newblock \bibinfo{booktitle}{\emph{Context-based topic models for query
  modification}}.
\newblock \bibinfo{type}{{T}echnical {R}eport}. \bibinfo{institution}{CIIR
  Technical Report, University of Massachusetts}.
\newblock


\bibitem[\protect\citeauthoryear{Hochreiter and Schmidhuber}{Hochreiter and
  Schmidhuber}{1997}]%
        {hochreiter1997long}
\bibfield{author}{\bibinfo{person}{Sepp Hochreiter} {and}
  \bibinfo{person}{J{\"u}rgen Schmidhuber}.} \bibinfo{year}{1997}\natexlab{}.
\newblock \showarticletitle{Long short-term memory}.
\newblock \bibinfo{journal}{\emph{Neural computation}} \bibinfo{volume}{9},
  \bibinfo{number}{8} (\bibinfo{year}{1997}), \bibinfo{pages}{1735--1780}.
\newblock


\bibitem[\protect\citeauthoryear{Kuzi, Carmel, Libov, and Raviv}{Kuzi
  et~al\mbox{.}}{2017}]%
        {kuzi2017query}
\bibfield{author}{\bibinfo{person}{Saar Kuzi}, \bibinfo{person}{David Carmel},
  \bibinfo{person}{Alex Libov}, {and} \bibinfo{person}{Ariel Raviv}.}
  \bibinfo{year}{2017}\natexlab{}.
\newblock \showarticletitle{Query expansion for email search}. In
  \bibinfo{booktitle}{\emph{Proceedings of the 40th International ACM SIGIR
  Conference on Research and Development in Information Retrieval}}.
  \bibinfo{pages}{849--852}.
\newblock


\bibitem[\protect\citeauthoryear{Li, Zhang, Bendersky, Deng, Metzler, and
  Najork}{Li et~al\mbox{.}}{2019}]%
        {li2019multi}
\bibfield{author}{\bibinfo{person}{Cheng Li}, \bibinfo{person}{Mingyang Zhang},
  \bibinfo{person}{Michael Bendersky}, \bibinfo{person}{Hongbo Deng},
  \bibinfo{person}{Donald Metzler}, {and} \bibinfo{person}{Marc Najork}.}
  \bibinfo{year}{2019}\natexlab{}.
\newblock \showarticletitle{Multi-view Embedding-based Synonyms for Email
  Search}. In \bibinfo{booktitle}{\emph{Proceedings of the 42nd International
  ACM SIGIR Conference on Research and Development in Information Retrieval}}.
  \bibinfo{pages}{575--584}.
\newblock


\bibitem[\protect\citeauthoryear{Liao, Jiang, Pei, Huang, Chen, Cao, and
  Li}{Liao et~al\mbox{.}}{2013}]%
        {liao2013vlhmm}
\bibfield{author}{\bibinfo{person}{Zhen Liao}, \bibinfo{person}{Daxin Jiang},
  \bibinfo{person}{Jian Pei}, \bibinfo{person}{Yalou Huang},
  \bibinfo{person}{Enhong Chen}, \bibinfo{person}{Huanhuan Cao}, {and}
  \bibinfo{person}{Hang Li}.} \bibinfo{year}{2013}\natexlab{}.
\newblock \showarticletitle{A vlHMM approach to context-aware search}.
\newblock \bibinfo{journal}{\emph{ACM Transactions on the Web (TWEB)}}
  \bibinfo{volume}{7}, \bibinfo{number}{4} (\bibinfo{year}{2013}),
  \bibinfo{pages}{1--38}.
\newblock


\bibitem[\protect\citeauthoryear{Lobel, Li, Gao, and Carin}{Lobel
  et~al\mbox{.}}{2019}]%
        {lobel2019ract}
\bibfield{author}{\bibinfo{person}{Sam Lobel}, \bibinfo{person}{Chunyuan Li},
  \bibinfo{person}{Jianfeng Gao}, {and} \bibinfo{person}{Lawrence Carin}.}
  \bibinfo{year}{2019}\natexlab{}.
\newblock \showarticletitle{RaCT: Toward Amortized Ranking-Critical Training
  For Collaborative Filtering}. In \bibinfo{booktitle}{\emph{International
  Conference on Learning Representations}}.
\newblock


\bibitem[\protect\citeauthoryear{Macdonald and Ounis}{Macdonald and
  Ounis}{2006}]%
        {macdonald2006combining}
\bibfield{author}{\bibinfo{person}{Craig Macdonald} {and} \bibinfo{person}{Iadh
  Ounis}.} \bibinfo{year}{2006}\natexlab{}.
\newblock \showarticletitle{Combining fields in known-item email search}. In
  \bibinfo{booktitle}{\emph{Proceedings of the 29th annual international ACM
  SIGIR conference on Research and development in information retrieval}}.
  \bibinfo{pages}{675--676}.
\newblock


\bibitem[\protect\citeauthoryear{Mackenzie, Gupta, Qiao, Awadallah, and
  Shokouhi}{Mackenzie et~al\mbox{.}}{2019}]%
        {mackenzie2019exploring}
\bibfield{author}{\bibinfo{person}{Joel Mackenzie}, \bibinfo{person}{Kshitiz
  Gupta}, \bibinfo{person}{Fang Qiao}, \bibinfo{person}{Ahmed~Hassan
  Awadallah}, {and} \bibinfo{person}{Milad Shokouhi}.}
  \bibinfo{year}{2019}\natexlab{}.
\newblock \showarticletitle{Exploring user behavior in email re-finding tasks}.
  In \bibinfo{booktitle}{\emph{The World Wide Web Conference}}.
  \bibinfo{pages}{1245--1255}.
\newblock


\bibitem[\protect\citeauthoryear{Matthijs and Radlinski}{Matthijs and
  Radlinski}{2011}]%
        {matthijs2011personalizing}
\bibfield{author}{\bibinfo{person}{Nicolaas Matthijs} {and}
  \bibinfo{person}{Filip Radlinski}.} \bibinfo{year}{2011}\natexlab{}.
\newblock \showarticletitle{Personalizing web search using long term browsing
  history}. In \bibinfo{booktitle}{\emph{Proceedings of the fourth ACM
  international conference on Web search and data mining}}.
  \bibinfo{pages}{25--34}.
\newblock


\bibitem[\protect\citeauthoryear{Meng, Karimzadehgan, Zhuang, and Metzler}{Meng
  et~al\mbox{.}}{2020}]%
        {meng2020separate}
\bibfield{author}{\bibinfo{person}{Yu Meng}, \bibinfo{person}{Maryam
  Karimzadehgan}, \bibinfo{person}{Honglei Zhuang}, {and}
  \bibinfo{person}{Donald Metzler}.} \bibinfo{year}{2020}\natexlab{}.
\newblock \showarticletitle{Separate and Attend in Personal Email Search}. In
  \bibinfo{booktitle}{\emph{Proceedings of the 13th International Conference on
  Web Search and Data Mining}}. \bibinfo{pages}{429--437}.
\newblock


\bibitem[\protect\citeauthoryear{Ogilvie and Callan}{Ogilvie and
  Callan}{2005}]%
        {ogilvie2005experiments}
\bibfield{author}{\bibinfo{person}{Paul Ogilvie} {and} \bibinfo{person}{Jamie
  Callan}.} \bibinfo{year}{2005}\natexlab{}.
\newblock \showarticletitle{Experiments with Language Models for Known-Item
  Finding of E-mail Messages.}. In \bibinfo{booktitle}{\emph{TREC}}.
\newblock


\bibitem[\protect\citeauthoryear{Qin, Li, Bendersky, and Metzler}{Qin
  et~al\mbox{.}}{2020}]%
        {qin2020matching}
\bibfield{author}{\bibinfo{person}{Zhen Qin}, \bibinfo{person}{Zhongliang Li},
  \bibinfo{person}{Michael Bendersky}, {and} \bibinfo{person}{Donald Metzler}.}
  \bibinfo{year}{2020}\natexlab{}.
\newblock \showarticletitle{Matching Cross Network for Learning to Rank in
  Personal Search}. In \bibinfo{booktitle}{\emph{Proceedings of The Web
  Conference 2020}}. \bibinfo{pages}{2835--2841}.
\newblock


\bibitem[\protect\citeauthoryear{Ramarao, Iyengar, Chitnis, Udupa, and
  Ashok}{Ramarao et~al\mbox{.}}{2016}]%
        {ramarao2016inlook}
\bibfield{author}{\bibinfo{person}{Pranav Ramarao}, \bibinfo{person}{Suresh
  Iyengar}, \bibinfo{person}{Pushkar Chitnis}, \bibinfo{person}{Raghavendra
  Udupa}, {and} \bibinfo{person}{Balasubramanyan Ashok}.}
  \bibinfo{year}{2016}\natexlab{}.
\newblock \showarticletitle{Inlook: Revisiting email search experience}. In
  \bibinfo{booktitle}{\emph{Proceedings of the 39th International ACM SIGIR
  conference on Research and Development in Information Retrieval}}.
  \bibinfo{pages}{1117--1120}.
\newblock


\bibitem[\protect\citeauthoryear{Robertson, Zaragoza, and Taylor}{Robertson
  et~al\mbox{.}}{2004}]%
        {robertson2004simple}
\bibfield{author}{\bibinfo{person}{Stephen Robertson}, \bibinfo{person}{Hugo
  Zaragoza}, {and} \bibinfo{person}{Michael Taylor}.}
  \bibinfo{year}{2004}\natexlab{}.
\newblock \showarticletitle{Simple BM25 extension to multiple weighted fields}.
  In \bibinfo{booktitle}{\emph{Proceedings of the thirteenth ACM international
  conference on Information and knowledge management}}.
  \bibinfo{pages}{42--49}.
\newblock


\bibitem[\protect\citeauthoryear{Shen, Karimzadehgan, Bendersky, Qin, and
  Metzler}{Shen et~al\mbox{.}}{2018}]%
        {shen2018multi}
\bibfield{author}{\bibinfo{person}{Jiaming Shen}, \bibinfo{person}{Maryam
  Karimzadehgan}, \bibinfo{person}{Michael Bendersky}, \bibinfo{person}{Zhen
  Qin}, {and} \bibinfo{person}{Donald Metzler}.}
  \bibinfo{year}{2018}\natexlab{}.
\newblock \showarticletitle{Multi-task learning for email search ranking with
  auxiliary query clustering}. In \bibinfo{booktitle}{\emph{Proceedings of the
  27th ACM International Conference on Information and Knowledge Management}}.
  \bibinfo{pages}{2127--2135}.
\newblock


\bibitem[\protect\citeauthoryear{Shen, Tan, and Zhai}{Shen
  et~al\mbox{.}}{2005}]%
        {shen2005implicit}
\bibfield{author}{\bibinfo{person}{Xuehua Shen}, \bibinfo{person}{Bin Tan},
  {and} \bibinfo{person}{ChengXiang Zhai}.} \bibinfo{year}{2005}\natexlab{}.
\newblock \showarticletitle{Implicit user modeling for personalized search}. In
  \bibinfo{booktitle}{\emph{Proceedings of the 14th ACM international
  conference on Information and knowledge management}}.
  \bibinfo{pages}{824--831}.
\newblock


\bibitem[\protect\citeauthoryear{Soboroff, de~Vries, and Craswell}{Soboroff
  et~al\mbox{.}}{2006}]%
        {soboroff2006overview}
\bibfield{author}{\bibinfo{person}{Ian Soboroff}, \bibinfo{person}{Arjen~P de
  Vries}, {and} \bibinfo{person}{Nick Craswell}.}
  \bibinfo{year}{2006}\natexlab{}.
\newblock \showarticletitle{Overview of the TREC 2006 Enterprise Track.}. In
  \bibinfo{booktitle}{\emph{Trec}}, Vol.~\bibinfo{volume}{6}.
  \bibinfo{pages}{1--20}.
\newblock


\bibitem[\protect\citeauthoryear{Tran, Karimzadehgan, Pasumarthi, Bendersky,
  and Metzler}{Tran et~al\mbox{.}}{2019}]%
        {tran2019domain}
\bibfield{author}{\bibinfo{person}{Brandon Tran}, \bibinfo{person}{Maryam
  Karimzadehgan}, \bibinfo{person}{Rama~Kumar Pasumarthi},
  \bibinfo{person}{Michael Bendersky}, {and} \bibinfo{person}{Donald Metzler}.}
  \bibinfo{year}{2019}\natexlab{}.
\newblock \showarticletitle{Domain Adaptation for Enterprise Email Search}. In
  \bibinfo{booktitle}{\emph{Proceedings of the 42nd International ACM SIGIR
  Conference on Research and Development in Information Retrieval}}.
  \bibinfo{pages}{25--34}.
\newblock


\bibitem[\protect\citeauthoryear{Ustinovskiy and Serdyukov}{Ustinovskiy and
  Serdyukov}{2013}]%
        {ustinovskiy2013personalization}
\bibfield{author}{\bibinfo{person}{Yury Ustinovskiy} {and}
  \bibinfo{person}{Pavel Serdyukov}.} \bibinfo{year}{2013}\natexlab{}.
\newblock \showarticletitle{Personalization of web-search using short-term
  browsing context}. In \bibinfo{booktitle}{\emph{Proceedings of the 22nd ACM
  international conference on Information \& Knowledge Management}}.
  \bibinfo{pages}{1979--1988}.
\newblock


\bibitem[\protect\citeauthoryear{van~der Maaten and Hinton}{van~der Maaten and
  Hinton}{2008}]%
        {Maaten08visualizingdata}
\bibfield{author}{\bibinfo{person}{Laurens van~der Maaten} {and}
  \bibinfo{person}{Geoffrey Hinton}.} \bibinfo{year}{2008}\natexlab{}.
\newblock \bibinfo{title}{Visualizing data using t-SNE}.
\newblock
\newblock


\bibitem[\protect\citeauthoryear{Vaswani, Shazeer, Parmar, Uszkoreit, Jones,
  Gomez, Kaiser, and Polosukhin}{Vaswani et~al\mbox{.}}{2017}]%
        {vaswani2017attention}
\bibfield{author}{\bibinfo{person}{Ashish Vaswani}, \bibinfo{person}{Noam
  Shazeer}, \bibinfo{person}{Niki Parmar}, \bibinfo{person}{Jakob Uszkoreit},
  \bibinfo{person}{Llion Jones}, \bibinfo{person}{Aidan~N Gomez},
  \bibinfo{person}{{\L}ukasz Kaiser}, {and} \bibinfo{person}{Illia
  Polosukhin}.} \bibinfo{year}{2017}\natexlab{}.
\newblock \showarticletitle{Attention is all you need}. In
  \bibinfo{booktitle}{\emph{Advances in neural information processing
  systems}}. \bibinfo{pages}{5998--6008}.
\newblock


\bibitem[\protect\citeauthoryear{Weerkamp, Balog, and De~Rijke}{Weerkamp
  et~al\mbox{.}}{2009}]%
        {weerkamp2009using}
\bibfield{author}{\bibinfo{person}{Wouter Weerkamp}, \bibinfo{person}{Krisztian
  Balog}, {and} \bibinfo{person}{Maarten De~Rijke}.}
  \bibinfo{year}{2009}\natexlab{}.
\newblock \showarticletitle{Using contextual information to improve search in
  email archives}. In \bibinfo{booktitle}{\emph{European Conference on
  Information Retrieval}}. Springer, \bibinfo{pages}{400--411}.
\newblock


\bibitem[\protect\citeauthoryear{White, Bennett, and Dumais}{White
  et~al\mbox{.}}{2010}]%
        {white2010predicting}
\bibfield{author}{\bibinfo{person}{Ryen~W White}, \bibinfo{person}{Paul~N
  Bennett}, {and} \bibinfo{person}{Susan~T Dumais}.}
  \bibinfo{year}{2010}\natexlab{}.
\newblock \showarticletitle{Predicting short-term interests using
  activity-based search context}. In \bibinfo{booktitle}{\emph{Proceedings of
  the 19th ACM international conference on Information and knowledge
  management}}. \bibinfo{pages}{1009--1018}.
\newblock


\bibitem[\protect\citeauthoryear{Xiang, Jiang, Pei, Sun, Chen, and Li}{Xiang
  et~al\mbox{.}}{2010}]%
        {xiang2010context}
\bibfield{author}{\bibinfo{person}{Biao Xiang}, \bibinfo{person}{Daxin Jiang},
  \bibinfo{person}{Jian Pei}, \bibinfo{person}{Xiaohui Sun},
  \bibinfo{person}{Enhong Chen}, {and} \bibinfo{person}{Hang Li}.}
  \bibinfo{year}{2010}\natexlab{}.
\newblock \showarticletitle{Context-aware ranking in web search}. In
  \bibinfo{booktitle}{\emph{Proceedings of the 33rd international ACM SIGIR
  conference on Research and development in information retrieval}}.
  \bibinfo{pages}{451--458}.
\newblock


\bibitem[\protect\citeauthoryear{Yahyaei and Monz}{Yahyaei and Monz}{2008}]%
        {yahyaei2008applying}
\bibfield{author}{\bibinfo{person}{Sirvan Yahyaei} {and}
  \bibinfo{person}{Christof Monz}.} \bibinfo{year}{2008}\natexlab{}.
\newblock \showarticletitle{Applying maximum entropy to known-item email
  retrieval}. In \bibinfo{booktitle}{\emph{European Conference on Information
  Retrieval}}. Springer, \bibinfo{pages}{406--413}.
\newblock


\bibitem[\protect\citeauthoryear{Zamani, Bendersky, Wang, and Zhang}{Zamani
  et~al\mbox{.}}{2017}]%
        {zamani2017situational}
\bibfield{author}{\bibinfo{person}{Hamed Zamani}, \bibinfo{person}{Michael
  Bendersky}, \bibinfo{person}{Xuanhui Wang}, {and} \bibinfo{person}{Mingyang
  Zhang}.} \bibinfo{year}{2017}\natexlab{}.
\newblock \showarticletitle{Situational context for ranking in personal
  search}. In \bibinfo{booktitle}{\emph{Proceedings of the 26th International
  Conference on World Wide Web}}. \bibinfo{pages}{1531--1540}.
\newblock


\end{thebibliography}

\end{document}